\documentclass{iopjournal}

\usepackage{enumitem} 
\usepackage{amssymb}
\usepackage{amsmath}
\usepackage{subcaption}
\usepackage{booktabs}
\usepackage{multirow}
\usepackage[numbers,square,sort&compress]{natbib}
\usepackage[T1]{fontenc}
\usepackage{lmodern}
\providecommand*{\Autoref}[1]{\begingroup
  \def\sectionautorefname{Section}%
  \autoref{#1}\endgroup}

\renewcommand{\sectionautorefname}{section}

\newlength{\labwidth}
\begin{document}

\articletype{Paper} 

\title{Robust Procurement Planning for Community Choice Aggregators: Managing Market Price Uncertainty}

\author{Manar Jaradat$^1$, James Himelic$^2$ and Neha Patankar$^{1,*}$}

\affil{$^1$School of Systems Science and Industrial Engineering, Binghamton University, Binghamton, NY, USA}

\affil{$^2$First Principles Advisory, Santa Cruz, CA, USA}

\affil{$^*$Author to whom any correspondence should be addressed.}

\email{npatankar@binghamton.edu}

\keywords{Robust optimization, Integrated resource planning, Community choice aggregators, Electricity market price uncertainty, Renewable energy procurement}

\begin{abstract}
Community Choice Aggregators (CCAs) play an increasingly important role in the energy transition by procuring renewable electricity on behalf of local communities while balancing affordability, resource adequacy, and clean-energy policy requirements. In California, these long-term procurement decisions are coordinated through integrated resource planning (IRP). However, conventional IRP frameworks typically represent wholesale electricity prices deterministically, even though market price uncertainty can expose CCAs to substantial financial risk by affecting the cost of market imports and, particularly, the value of export opportunities. This paper develops a robust optimization framework for CCA resource procurement that explicitly hedges against uncertainty in wholesale import and export prices. We extend the open-source GenX capacity expansion model with a budgeted robust optimization formulation that protects procurement decisions against adverse price realizations while remaining computationally tractable for large-scale, multistage planning studies. Using data from a California CCA, we show that accounting for market price uncertainty can materially change optimal procurement strategies. Rather than aggressively expanding renewable capacity based on expected export revenues, robust portfolios reduce dependence on uncertain future market revenues, leading to more conservative renewable investments and less reliance on market exports. We further demonstrate that moderate uncertainty budgets provide most of the risk-reduction benefits of robust planning at a limited additional cost, whereas higher robustness levels yield diminishing benefits as portfolio conservatism increases. These findings provide practical guidance for selecting uncertainty budgets in IRP studies and demonstrate how robust optimization can support affordable and resilient procurement decisions as CCAs contribute to the transition toward low-carbon electricity systems.
\end{abstract}


\section{INTRODUCTION} 
\label{sec:introduction}
Community Choice Aggregators (CCAs) have become important actors in the electricity transition because they allow local governments to procure power on behalf of retail customers and choose resource mixes that reflect community priorities, including cleaner electricity, while investor-owned utilities continue to own and operate transmission and distribution infrastructure \citep{o2019empowered, Hess2020Energy}. First introduced in Massachusetts in 1997 as part of electric sector restructuring \citep{Hsu2022Straight}, CCAs have expanded through policy diffusion and local adoption. As of 2022, more than 1,800 local governments in six states served over 36 million customers through CCAs \citep{Hsu2022Straight}. In California, CCAs were enabled by Assembly Bill 117 following the 2000–2001 energy crisis, to increase local control over electricity procurement. Twenty-five operational CCAs serve more than 14 million customers statewide \citep{calcca_powered}. Between 2011 and 2019, California CCAs purchased 23.5 million MWh of renewable electricity beyond state-mandated levels. When required and voluntary purchases are combined, they procured more than twice their collective baseline state requirement \citep{Trumbull2021The}. Their long-term procurement decisions can therefore influence the scale and timing of clean-energy investment, as well as the cost of the transition for their customers.

Like other load-serving entities (LSEs), CCAs must submit Integrated Resource Plans (IRPs) that identify reliable, cost-effective resource portfolios capable of meeting forecast demand and applicable policy requirements. In California, IRPs must align with the Renewable Portfolio Standard, greenhouse-gas reduction targets established under Senate Bill 350 \citep{SB350_2015}, and the statutory requirement for 100\% clean electricity by 2045 \citep{CA_SB100_2018, CPUC_RSP_2022}. Demand forecasts are provided by the California Energy Commission and refined by individual LSEs to reflect conditions in their service areas. LSEs then evaluate candidate portfolios and submit preferred plans to the California Public Utilities Commission (CPUC) for review \citep{Carvallo2016Load}. For a CCA, this process requires decisions about how much renewable generation and storage to procure, when to make those commitments, and how much to rely on wholesale market transactions over the planning horizon.

Wholesale markets affect both sides of that decision. CCAs may purchase electricity when their contracted and owned resources do not cover demand and sell surplus generation when output exceeds their needs. California Independent System Operator (CAISO) market prices consequently affect the cost of serving customers and the expected revenue from additional generation. This second effect matters for clean-energy investment: a resource that produces electricity beyond the CCA’s own requirements may appear economical because the planning model assigns value to future market sales. Yet the capacity commitment is long-lived, while the revenue ultimately earned from exports depends on uncertain future prices. An unfavorable change in those prices can leave the CCA with a portfolio that still meets its clean-energy obligations but costs more than anticipated.

Market transactions also need to be distinguished from dependable capacity in long-term planning. CAISO operates the transmission network and administers the wholesale market, dispatching in-state resources and participating out-of-state generators to meet demand. Some external resources participate through dynamic scheduling, but market-driven imports do not necessarily provide firm procurement commitments that a CCA can count toward its long-term resource needs \citep{CPUC_IRP_Inputs_2025}. CCAs must therefore plan for reliability and policy compliance while assessing the financial consequences of buying and selling electricity in a market whose future prices cannot be known when investments are made.

This creates a planning challenge for the energy transition. Deterministic IRP tools commonly evaluate portfolios using specified wholesale price trajectories. Those prices influence which clean resources are selected and whether building capacity beyond policy requirements appears profitable. If projected export revenues are treated as dependable, a deterministic plan may favor additional renewable investment whose economic case weakens under adverse market prices. Conversely, ignoring the possibility of costly imports can understate the expense of meeting demand with market purchases. CCAs face an added institutional concern: they must maintain competitive retail rates to limit customer opt-outs and may have less financial capacity than large investor-owned utilities to absorb prolonged departures from planned procurement costs. A clean-energy portfolio must therefore be assessed for both policy compliance and its exposure to uncertain market prices.

IRP decisions depend on many uncertain inputs, including investment and fixed O\&M costs, demand growth, renewable resource availability, and wholesale market prices \cite{Goteti2025Stochastic, dong2023hybrid, mohseni2022demand}. Previous studies have extensively examined uncertainty in capital costs \cite{patankar2022using, moret2020decision}, electricity demand \cite{mejia2014adjustable, wu2024distributionally}, and renewable generation \cite{abdin2022optimizing, wang2023two, xu2015risk}. Wholesale price uncertainty has received comparatively limited attention in long-term CCA procurement, particularly its effect on the value of surplus clean generation. That omission matters because market prices can influence investment decisions even when emissions limits and clean-energy procurement requirements are held fixed.

Several optimization approaches can represent uncertainty in energy planning. Stochastic programming uses probabilistic scenarios and commonly minimizes expected cost \cite{Yu2024On, Rathi2022Capacity, Goteti2025Stochastic}. It is useful when probability distributions can be characterized, but large-scale, multistage planning problems can require decomposition techniques to remain tractable \cite{bertsimas2022robust, ZUGNO2015659}. Scenario-based methods do not necessarily require explicit probability distributions, although their conclusions depend on the selection and coverage of the scenarios examined \cite{pourahmadi2019distributionally, Hu2022Scenario}. Robust optimization (RO) offers a complementary approach by protecting decisions against adverse realizations within a specified uncertainty set. Classical RO can produce overly conservative solutions; the budget-of-uncertainty framework of Bertsimas and Sim \cite{bertsimas2004price} allows decision-makers to control the degree of protection. Distributionally robust optimization addresses uncertainty about the underlying probability distribution, generally with additional computational requirements \citep{Alismail2018DRO}.

RO formulations can be static or adaptive. Adaptive RO allows decisions to change after uncertainty is revealed, but can substantially increase the complexity of multistage capacity expansion models. In the CCA planning context studied here, long-term contracts and capacity investments cannot readily be reversed in response to each realization of future market prices. We therefore use a budgeted static RO formulation to select procurement portfolios that account for adverse wholesale prices without making investment decisions contingent on a particular price realization. Investments occur across multiple planning stages, but decisions within each stage are not adjusted after uncertain prices are realized. This structure supports extensive sensitivity analysis while reflecting the long-term nature of procurement commitments.

Widely used capacity expansion tools offer limited native support for this type of analysis. RESOLVE, used in California’s IRP process, applies a deterministic planning formulation \citep{e3_resolve}. Open-source tools such as PyPSA \citep{PyPSA} and Temoa support scenario-based or stochastic analyses \citep{pampado2025exploring, Fioriti2025}, while GridPath and OpenTEPES represent stochastic or operational uncertainty in specific applications. To the best of our knowledge, no widely used open-source capacity expansion model provides a budgeted RO framework designed to examine how uncertain wholesale import costs and export revenues affect long-term CCA procurement.

This study asks how explicitly accounting for wholesale price uncertainty changes a CCA’s clean-energy investment strategy and financial exposure. We extend the open-source GenX capacity expansion model \citep{genx} to represent key features of California CCA planning and implement a budgeted static RO formulation for market prices. We then examine how the uncertainty budget changes the amount, timing, and composition of renewable and storage procurement; distinguish changes in market imports from changes in exports and their expected revenues; and evaluate the resulting portfolios under out-of-sample price trajectories. Finally, we compare planning cost, downside cost exposure relative to each portfolio’s nominal cost, portfolio stability, and dependence on market imports.

The central contribution is to show how price risk can change the \textbf{path} a CCA takes to meet clean-energy requirements even when those requirements remain fixed. In the case studied, accounting for adverse market prices reduces the incentive to build renewable capacity whose economic value depends on uncertain export revenues. Robust portfolios continue to satisfy the modeled clean-energy and emissions constraints, but differ in their investments, market transactions, and financial exposure. This distinction gives CCAs a way to evaluate how much protection against market price risk they want to incorporate into long-term procurement, and what that protection means for the affordability and durability of their clean-energy transition.

The remainder of the paper is organized as follows. \Autoref{sec:model_formulation} introduces the baseline capacity expansion model implemented in GenX and describes the enhancements required to align it with the California IRP process. \Autoref{sec:case_study} presents the case study and details the construction of the uncertainty set used in the analysis. \Autoref{sec:results} reports results from the deterministic model and examines the impact of uncertainty through sensitivity analyses with respect to the budget of uncertainty and the specification of the uncertainty set. Finally, \autoref{sec:conclusion} concludes the paper and discusses implications for CCA procurement planning.

\section{Model Formulation}
\label{sec:model_formulation}
We extend the open-source GenX capacity expansion model \citep{genx} to represent key features of CCA integrated resource planning and to account for uncertainty in wholesale market prices. The underlying model minimizes the net present value (NPV) of investment and operating costs over a multistage planning horizon, subject to resource investment, dispatch, power-balance, policy, and resource-adequacy constraints. To align GenX with the California IRP process, we incorporate an annual planning reserve margin (PRM) requirement based on resource-specific capacity accreditation, a market module representing unspecified imports and exports, and emissions accounting for imported electricity. The complete extended GenX formulation and corresponding nomenclature are provided in Appendix~\ref{app:extended_genx}. The following subsections briefly describe the wholesale market representation and present the robust optimization formulation that is central to this study.

\subsection{Wholesale Market Representation}
\label{ssec:market}

Unspecified imports are represented through wholesale market transactions. Electricity imported into zone $z$ during time step $t$, denoted by $\Pi^{\text{Mkt}}_{z,t}$, is purchased at price $\pi^{\text{Mkt,Imp}}_{z,t}$, while exported electricity $\Theta^{\text{Mkt}}_{z,t}$ is sold at price $\pi^{\text{Mkt,Exp}}_{z,t}$. Their contribution to the objective function is

\begin{equation}
\sum_{z \in M}\sum_{t \in T}
\omega_t
\left(
\pi^{\text{Mkt,Imp}}_{z,t}\Pi^{\text{Mkt}}_{z,t}
-
\pi^{\text{Mkt,Exp}}_{z,t}\Theta^{\text{Mkt}}_{z,t}
\right).
\label{eq:market_obj}
\end{equation}

where $\omega_t$ is the weight assigned to time step $t$ when calculating annual operating costs.

Imports and exports are incorporated into the zonal power balance, subject to transaction limits and constraints preventing unsupported exports. Emissions associated with unspecified imports are also included in the zonal emissions balance to maintain consistency with California IRP accounting requirements. The complete market formulation is provided in Appendix~\ref{app:extended_genx}.

\subsection{Robust Optimization Model}
\label{ssec:ro}

Although long-term planning is affected by uncertainty in demand, technology costs, fuel prices, and renewable availability, these sources of uncertainty have been widely studied within stochastic and robust planning frameworks. In contrast, uncertainty in wholesale market prices has received comparatively limited attention despite its growing importance for CCAs, whose procurement strategies increasingly rely on non-firm market transactions. Accordingly, this study intentionally isolates market price uncertainty to examine its independent impact on procurement decisions. 

To capture uncertainty in volatile market import and export prices, the deterministic capacity expansion model is extended using RO framework that hedges procurement decisions against adverse price realizations. Price uncertainty is represented through one-sided deviations in market import prices, $\Delta\pi^{\text{Mkt,Imp}}_{z,t}$, and market export prices, $\Delta\pi^{\text{Mkt,Exp}}_{z,t}$, which affect the cost of market imports $\Pi^{\text{Mkt}}_{z,t}$ and the revenue from market exports $\Theta^{\text{Mkt}}_{z,t}$ introduced in the market module. 

RO provides a conservative, distribution-free approach that protects the planning solution against worst-case price movements without requiring probabilistic assumptions. Because not all prices are expected to attain their worst-case deviations simultaneously, uncertainty is structured using the budget-of-uncertainty framework of \citep{bertsimas2004price}, where the scalar $\Gamma$ denotes the uncertainty budget and limits the number of price deviations that can be jointly active.

The resulting problem is formulated as a budgeted static robust optimization model with an inner adversarial problem. The planner selects the investment, operational, and market-transaction decisions, including resource capacities $\Omega_y$, market imports $\Pi^{\text{Mkt}}_{z,t}$, and market exports $\Theta^{\text{Mkt}}_{z,t}$. Given these decisions, an adversarial decision-maker, referred to as ``nature,'' selects the combination of adverse market price deviations that maximizes the additional system cost within the specified uncertainty budget. These deviations are represented by the activation variables $s^{\text{Mkt,Imp}}_{z,t}$ and $s^{\text{Mkt,Exp}}_{z,t}$, which indicate the fraction of the maximum adverse deviation applied to each import and export price. The resulting robust objective is given by

\begin{subequations}
\label{eq:ro_primal}

\begin{equation}
\label{eq:ro_primal_obj}
\begin{aligned}
\min\;\text{NPV}
+
\max_{s} \sum_{z \in M} \sum_{t \in T}
\omega_t \Big(
\Delta\pi^{\text{Mkt,Imp}}_{z,t}\,\Pi^{\text{Mkt}}_{z,t}\,s^{\text{Mkt,Imp}}_{z,t}
+
\Delta\pi^{\text{Mkt,Exp}}_{z,t}\,\Theta^{\text{Mkt}}_{z,t}\,s^{\text{Mkt,Exp}}_{z,t}
\Big)
\end{aligned}
\end{equation}

\text{s.t.}
\begin{align}
0 \le s^{\text{Mkt,Imp}}_{z,t} &\le 1,
\quad \forall z \in M,\; t \in T,
\quad (q^{\text{Mkt,Imp}}_{z,t})
\\[3pt]
0 \le s^{\text{Mkt,Exp}}_{z,t} &\le 1,
\quad \forall z \in M,\; t \in T,
\quad (q^{\text{Mkt,Exp}}_{z,t})
\\[3pt]
\sum_{z,t}
\left(
s^{\text{Mkt,Imp}}_{z,t} +
s^{\text{Mkt,Exp}}_{z,t}
\right)
&\le \Gamma,
\quad (p)
\end{align}

\end{subequations}

The uncertainty budget $\Gamma$ controls the conservativeness of the robust solution by limiting the total number of price deviations that may simultaneously attain their worst-case values. Although $\Gamma$ is a numerical budget, we report it as a percentage of the total number of uncertain price parameters to simplify interpretation across planning stages. Thus, a reported value of $\Gamma=8\%$, for example, represents a numerical budget equal to $8\%$ of the uncertain parameters rather than a percentage-valued model parameter. A value of $\Gamma=0$ reproduces the deterministic solution, in which no price deviations are allowed. As $\Gamma$ increases, a larger fraction of price deviations may become active. In the evaluated solutions, imports and exports do not occur simultaneously within the same time step; consequently, a budget equal to $50\%$ of the uncertain parameters represents the most conservative effective setting. Beyond this threshold, additional budget cannot activate further adverse deviations associated with realized market transactions.

In its current form, the resulting min--max formulation is not directly tractable using standard optimization solvers due to the nested maximization over uncertain price deviations. To obtain a computationally tractable model, we apply strong duality. In linear programming, strong duality guarantees that, under mild feasibility conditions, the optimal value of a primal problem equals that of its dual. This property allows the inner worst-case price-deviation problem to be replaced exactly by its dual minimization problem, eliminating the nested max operator and yielding a single-level linear program that can be solved efficiently.

By dualizing the inner worst-case price-deviation problem and introducing the corresponding dual variables, $p$ associated with the uncertainty budget constraint and $q^{\text{Mkt,Imp}}_{z,t}$ and $q^{\text{Mkt,Exp}}_{z,t}$ associated with the deviation bounds, we obtain the following dual formulation of the inner problem:

\begin{subequations}
\label{eq:ro_dual_system}

\begin{equation}
\label{eq:ro_dual_obj}
\begin{aligned}
\min_{p,\, q}\;
\sum_{z,t} q^{\text{Mkt,Imp}}_{z,t}
+
\sum_{z,t} q^{\text{Mkt,Exp}}_{z,t}
+
\Gamma \cdot p
\end{aligned}
\end{equation}

\begin{align}
q^{\text{Mkt,Imp}}_{z,t} + p
&\ge \omega_t \, \Delta\pi^{\text{Mkt,Imp}}_{z,t} \, \Pi^{\text{Mkt}}_{z,t},
&\forall z \in M,\; t \in T,
\label{eq:ro_dual_imp}
\\[3pt]
q^{\text{Mkt,Exp}}_{z,t} + p
&\ge \omega_t \, \Delta\pi^{\text{Mkt,Exp}}_{z,t} \, \Theta^{\text{Mkt}}_{z,t},
&\forall z \in M,\; t \in T.
\label{eq:ro_dual_exp}
\end{align}

\end{subequations}

Substituting the dual formulation in \eqref{eq:ro_dual_system} into the master problem yields the following equivalent linear robust objective:

\begin{equation}
\label{eq:ro_final}
\min \;
\text{NPV}
+
\sum_{z,t} q^{\text{Mkt,Imp}}_{z,t}
+
\sum_{z,t} q^{\text{Mkt,Exp}}_{z,t}
+
\Gamma \cdot p.
\end{equation}

This formulation produces a resource portfolio that is protected against adverse market price realizations within the specified uncertainty set. Because the investment, operational, and market-transaction decisions do not adapt after uncertain prices are realized, the formulation is static and may produce more conservative solutions than an adaptive approach. Although the capacity expansion model spans multiple planning stages, these stages represent the timing of investment decisions rather than adaptation to observed market prices. This structure provides a transparent and computationally tractable framework for evaluating the cost and procurement implications of market price uncertainty in IRP applications.

\section{Case Study}
\label{sec:case_study}

The case study was developed in collaboration with a CCA operating in California using data and planning inputs provided by the organization. These inputs include the CCA's existing and candidate resource portfolios, investment and operating cost assumptions, market price forecasts, load projections, and renewable resource availability over the planning horizon. The analysis covers the period from 2025 to 2040. To reduce the model size while retaining greater temporal detail in the near term, the planning horizon is represented by five periods beginning in 2025, 2026, 2028, 2030, and 2035, with corresponding lengths of 1, 2, 2, 5, and 5 years. This structure uses coarser temporal resolution in later years, when planning inputs are inherently more uncertain. The existing resource portfolio at the beginning of 2025 is inherited from the CCA's 2024 portfolio, after which investment and operational decisions evolve over the study horizon. This temporal aggregation is consistent with the multistage representation commonly used in long-term IRP studies. Model assumptions and policy parameters are aligned with the CPUC IRP Inputs and Assumptions document \citep{CPUC_IRP_Inputs_2025}, which defines the standard datasets, modeling conventions, and regulatory requirements applicable to each IRP cycle.

\subsection{Capacity Expansion Model Inputs}
\label{subsec:cem_inputs}
The capacity expansion analysis uses the resource portfolio provided by the collaborating CCA. The portfolio includes a diverse set of technologies, including thermal generators, energy storage, variable renewable resources, must-run resources, and hybrid resources, which combine a renewable generator and an energy storage system at the same site and may share interconnection infrastructure, each characterized by detailed operational specifications. While the optimization model represents all 65 resources individually, results are reported in aggregated technology categories.

The portfolio includes both dispatchable and non-dispatchable resources that contribute toward meeting the PRM. Consistent with CPUC IRP practice, the demand response (DR) categories \texttt{CAISO\_Conventional\_DR} and \texttt{Conventional\_DR} represent load-modifying or RA-eligible demand response programs rather than energy-producing resources. Because these technologies reduce demand during peak events rather than generate electricity \citep{CPUC_IRP_Inputs_2025}, they are modeled as thermal-type resources with intentionally high heat rates (20 MMBTU/MWh) and prohibitively high variable operating costs (1000 \$/MWh). This formulation prevents dispatch in the hourly energy balance while still allowing these resources to contribute firm capacity toward PRM requirements. The \texttt{RA\_Only} category is treated similarly, representing capacity that contributes to resource adequacy obligations but does not participate in system dispatch.

Although GenX allows hybrid resources to charge from the grid by default, this behavior is controlled in our implementation through an input-based switch to reflect IRP eligibility rules. Grid charging can be turned on or off at the resource level based on procurement and clean-energy accounting constraints.

\begin{figure}[htbp]
  \centering
  \subfloat[]{
     \includegraphics[width=0.6\linewidth]{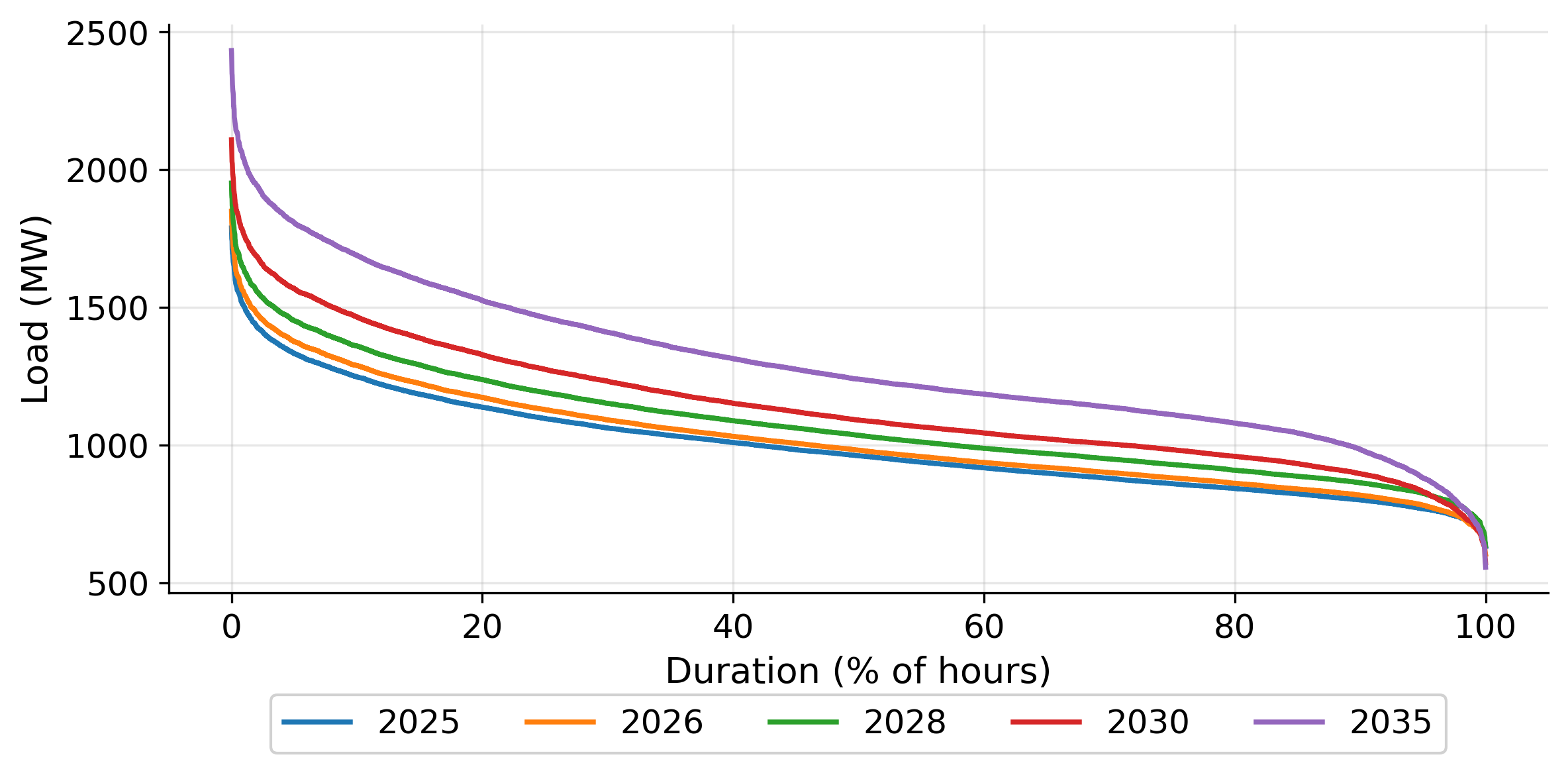}
    \label{fig:load_ldc}
  }
  \hfill
  \subfloat[]{
    \includegraphics[width=0.6\linewidth]{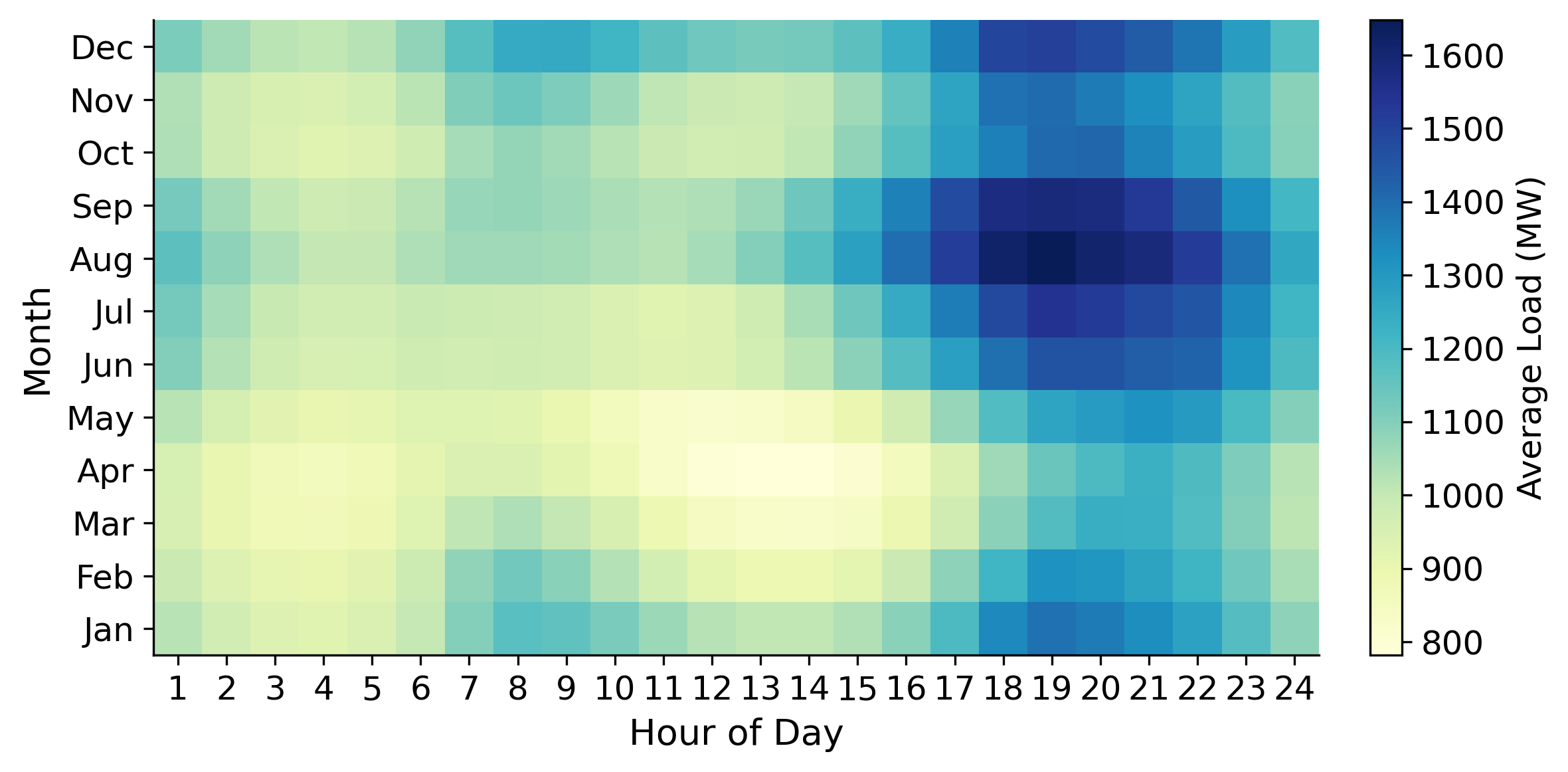}
    \label{fig:load_heatmap}
  }
  \caption{Electricity load characteristics across the planning horizon. 
(a) Load duration curves for each planning year. 
(b) Average hourly load by month and hour of day.}
  \label{fig:load_inputs}
\end{figure}

The load duration curves in \autoref{fig:load_ldc} summarize the hourly demand distribution across the five planning stages, illustrating the growth in total demand and the widening of peak hours over time. The heatmap of average load by month and hour in \autoref{fig:load_heatmap} highlights the clear temporal structure of demand: diurnal peaks appear in the late afternoon and early evening, while midday loads remain lower due to behind-the-meter and utility-scale solar generation. Clear seasonal patterns emerge, with energy demand increasing during the summer and early autumn. These temporal characteristics determine resource adequacy requirements and guide investment in both firm and flexible energy sources. They also influence exposure to market prices, particularly during peak hours, and thus shape the role of hedging strategies in managing uncertainty.

According to the CPUC Inputs and Assumptions framework \citep{CPUC_IRP_Inputs_2025}, LSEs must comply with statewide requirements governing emissions limits, renewable and clean-energy procurement, and system reliability. For this case study, the CCA’s proportional obligations translate into a declining CO\textsubscript{2} cap, increasing clean-energy shares under the Energy Share Requirements (ESR), and a set of reliability-driven limits on both annual and cumulative build capacity for solar, wind, and energy storage technologies.
These requirements, summarized in \autoref{tab:policy_requirements}, ensure consistency with the IRP process and define the feasible pathway for resource development across the planning stages. Together, the declining emissions cap and increasing clean-energy requirement represent the CCA's pathway toward a progressively lower-carbon resource portfolio over the planning horizon. Because GenX inherently imposes technology buildout constraints cumulatively over the entire planning horizon, we introduce stage-specific construction constraints to ensure that these limits are enforced within each planning period rather than only at the end of the horizon.

\begin{table}[h!]
\centering
\scriptsize
\caption{Policy and resource development constraints applied across the five planning years}
\label{tab:policy_requirements}
\resizebox{0.9\linewidth}{!}{
\begin{tabular}{lccccc}
\hline
\textbf{Constraint} & \textbf{2025} & \textbf{2026} & \textbf{2028} & \textbf{2030} & \textbf{2035} \\
\hline
$\text{CO}_{2}$ cap (Mtons) & 2.368 & 2.158 & 1.772 & 1.362 & 1.055 \\
ESR requirement (fraction) & 0.245 & 0.331 & 0.514 & 0.719 & 0.793 \\
\hline
\multicolumn{6}{l}{\textbf{Energy storage (BESS) constraints}} \\
Max annual BESS build (MW) & 200 & 350 & 500 & 500 & 500 \\
Max BESS installed capacity (MW) & 350 & 700 & 1550 & 2500 & 2500 \\
\hline
\multicolumn{6}{l}{\textbf{Solar PV constraints}} \\
Max annual solar build (MW) & 150 & 500 & 500 & 500 & 500 \\
Max solar installed capacity (MW) & 300 & 800 & 1800 & 2500 & 2500 \\
\hline
\multicolumn{6}{l}{\textbf{Wind constraints}} \\
Max annual wind build (MW) & 100 & 300 & 500 & 500 & 500 \\
Max wind installed capacity (MW) & 250 & 550 & 1350 & 2500 & 2500 \\
\hline
PRM requirement (MW) & 1468 & 1525 & 1430 & 1504 & 1314 \\
\hline
\end{tabular}
}
\end{table}

\subsection{Market Price Inputs and Uncertainty Set Construction}
\label{subsec:uncertainty_set_construction}
Six hourly market import price trajectories were provided by the collaborating CCA for the 2025--2040 planning horizon. These trajectories represent alternative long-term price projections used by the CCA for resource planning. The underlying baseline forecast was not shared. Instead, the CCA provided six scenarios generated by applying normally distributed perturbations to its proprietary forecast: $P_{\text{base\_high}}$, $P_{\text{base\_low}}$, $P_{\text{low\_low}}$, $P_{\text{high\_high}}$, $P_{\text{base\_base}}$, and $P_{\text{base\_flat}}$.

\begin{figure}[htbp]
  \centering
  \subfloat[]{
     \includegraphics[width=0.46\linewidth]{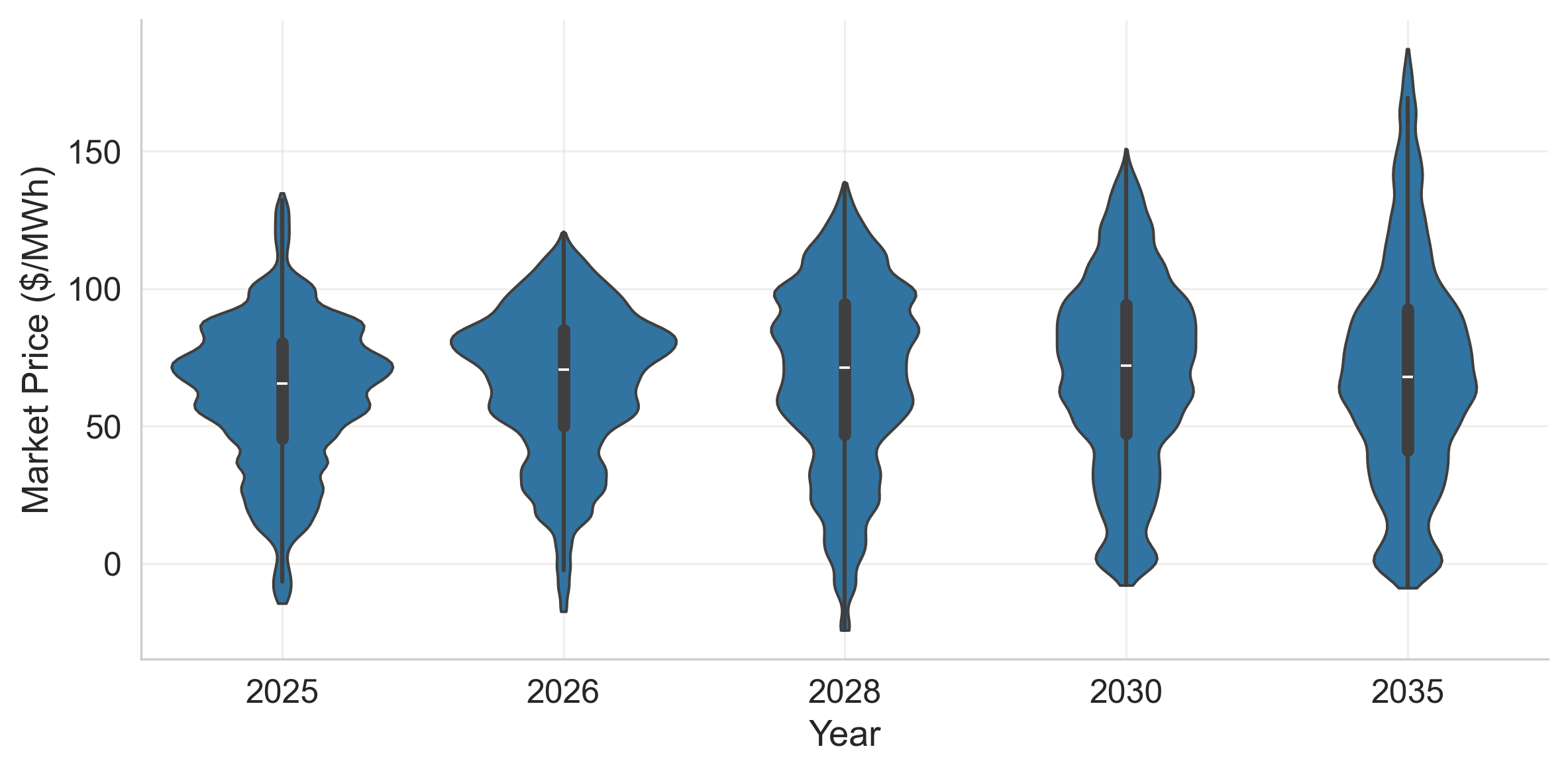}
    \label{fig:prices_violin}
  }
  \hfill
    \subfloat[]{
    \includegraphics[width=0.46\linewidth]{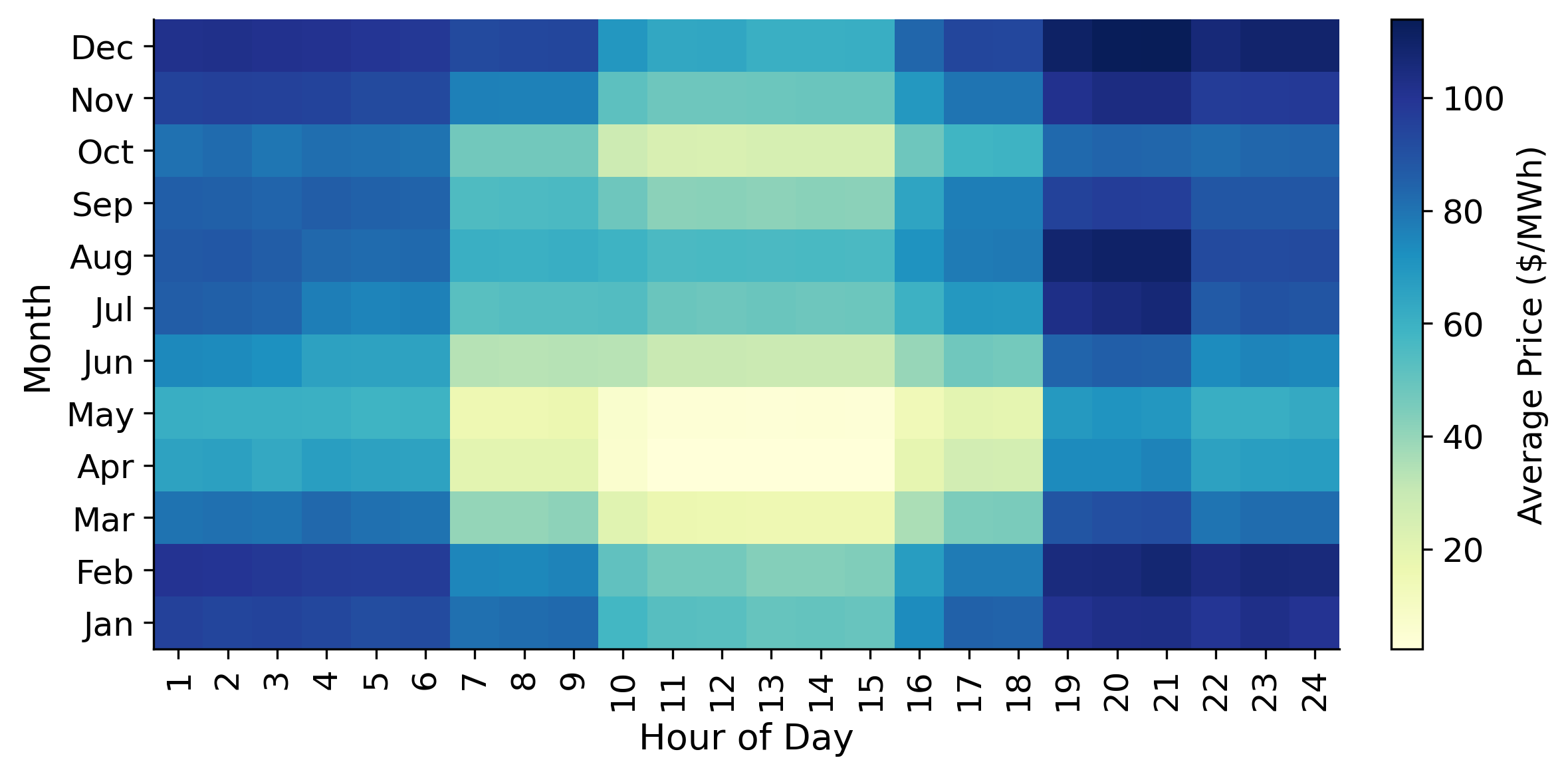}
    \label{fig:price_heatmap}
  }  
  \caption{Wholesale market import price characteristics.
  (a) Distribution of hourly market prices across the six scenarios for selected planning years.
  (b) Average market price by month and hour of day across the selected planning years.}
  \label{fig:price_inputs}
\end{figure}

The trajectories retain strong diurnal and seasonal patterns while diverging more substantially in later planning years. As shown in \autoref{fig:prices_violin}, the distribution of hourly prices widens over the planning horizon. \autoref{fig:price_heatmap} further shows that average prices are generally lower during midday hours, when solar generation is abundant, and higher during evening hours as solar output declines and net load increases. This widening spread reflects increasing uncertainty over the planning horizon and provides the basis for constructing the market price uncertainty set.

To preserve these temporal characteristics, including diurnal patterns, seasonal variation, and negative price signals, we select one of the observed trajectories as the representative nominal profile rather than using a smoothed hourly average or median series. For each planning year, the mean absolute deviation (MAD) of each trajectory from the ensemble mean was calculated as a diagnostic measure of representativeness. Based on this analysis, $P_{\text{base\_base}}$ was selected as the nominal import price trajectory $\pi^{\text{Mkt,Imp}}_{t}$ because it consistently exhibited the lowest deviation from the ensemble during the later planning years while preserving realistic hour-to-hour price dynamics.

The uncertainty set is constructed as an asymmetric interval around the nominal trajectory, defining the range of adverse price realizations against which procurement decisions are protected. The nominal export price is set to $0.95\,\pi^{\text{Mkt,Imp}}_{t}$, creating a 5\% difference between import and export prices that reflects the distinct values of market purchases and sales and discourages simultaneous market transactions.

An important feature of the CCA-provided scenarios is that $P_{\text{high\_high}}$ does not form a pointwise upper bound on the nominal trajectory. During some hours, its price is lower than the corresponding nominal price because each scenario represents an internally consistent long-term price trajectory rather than an hour-by-hour bound. Absolute differences from the nominal trajectory are therefore used to construct consistently adverse deviations for the robust optimization model.

For market imports, the upward deviation $\Delta\pi^{\text{Mkt,Imp}}_{t}$ is defined as the absolute difference between the nominal import price and the corresponding price in $P_{\text{high\_high}}$. For market exports, the downward deviation $\Delta\pi^{\text{Mkt,Exp}}_{t}$ is defined as the absolute difference between the nominal export price and the corresponding export price derived from $P_{\text{low\_low}}$. This construction preserves the hour-specific structure of the price scenarios while ensuring that import-price deviations increase procurement costs and export-price deviations reduce market revenues relative to their nominal values.

Let $\widetilde{\pi}^{\text{Mkt,Imp}}_{t}$ and $\widetilde{\pi}^{\text{Mkt,Exp}}_{t}$ denote the realized import and export prices, respectively. The resulting uncertainty bounds are

\begin{align*}
\pi^{\text{Mkt,Imp}}_{t}
&\le
\widetilde{\pi}^{\text{Mkt,Imp}}_{t}
\le
\pi^{\text{Mkt,Imp}}_{t}
+
\Delta\pi^{\text{Mkt,Imp}}_{t}, \\
0.95\,\pi^{\text{Mkt,Imp}}_{t}
-
\Delta\pi^{\text{Mkt,Exp}}_{t}
&\le
\widetilde{\pi}^{\text{Mkt,Exp}}_{t}
\le
0.95\,\pi^{\text{Mkt,Imp}}_{t}.
\end{align*}

\begin{figure}[htbp]
  \centering
  \includegraphics[width=1.0\textwidth]{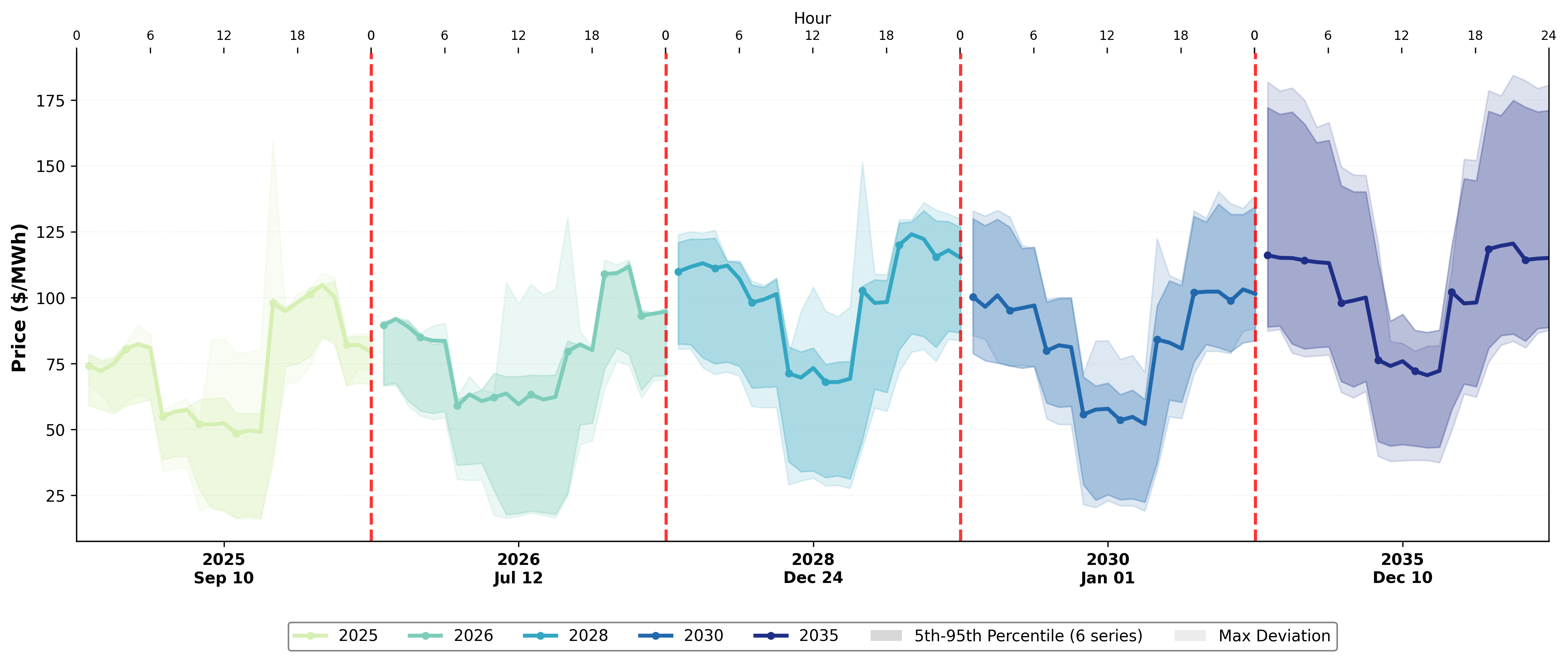}
  \caption{Hourly market import price profiles for representative high-price days across planning years (2025--2035). For each year, the solid line shows the nominal price trajectory ($P_{\text{base\_base}}$). The darker shaded band denotes the empirical 5th–95th percentile range across the six vendor-provided price trajectories at each hour. The lighter shaded band represents the asymmetric uncertainty bounds constructed from absolute deviations relative to the nominal profile and used to define the uncertainty set.}

  \label{fig:uncertainty_set_representative_days}
\end{figure}

\autoref{fig:uncertainty_set_representative_days} illustrates how market price uncertainty evolves over time by showing a representative high-price day from each planning year. The width of the constructed uncertainty envelope increases over the planning horizon. For market imports, the average deviation from the nominal $P_{\text{base\_base}}$ trajectory grows from approximately \$8/MWh in 2025 to more than \$34/MWh by 2035. For market exports, the average deviation from the corresponding nominal export price trajectory rises from approximately \$20/MWh in 2025 to over \$29/MWh by 2035.

Across the planning years, the largest deviations generally occur during the evening ramp hours, approximately 18:00--22:00, when nominal market prices are also elevated, as shown in the hourly price heatmap in \autoref{fig:price_heatmap}. By representing these hourly variations directly rather than relying on annual or average uncertainty measures, the uncertainty set preserves the periods of concentrated price dispersion that may have the greatest influence on procurement and market decisions.

\subsection{Simulation and Computational Setup}

All scenarios were solved in full-foresight, multistage mode using the extended GenX capacity expansion model. Each scenario was evaluated across a range of robustness levels, with the uncertainty budget $\Gamma$ ranging from 0, corresponding to the deterministic case, to a numerical budget equal to 50\% of the uncertain price parameters, corresponding to the most conservative effective setting. This systematic sweep enables a consistent comparison of how portfolio composition, market transactions, and total system cost respond to increasing protection against price uncertainty under alternative price structures and export assumptions.

The multistage GenX implementation uses Plasmo.jl to organize planning stages and the capacity-linking constraints between them within a graph-structured modeling framework built on JuMP. Although the extended model includes optional binary constraints that can prevent simultaneous market imports and exports, these constraints were not enforced in the experiments presented here. The combination of market price assumptions and export restrictions was sufficient to prevent simultaneous transactions in the resulting solutions. Consequently, all scenarios were solved as linear programs.

All optimization experiments were conducted on Binghamton University's Spiedie high-performance computing cluster \citep{spiedie_hpc}. Each job was allocated five CPU cores and 32~GB of RAM. The extended GenX model was implemented in Julia~1.10.4 and solved using Gurobi~12.0.3.

\section{Results}
\label{sec:results}

\subsection{Baseline Procurement under Deterministic Prices}
\label{subsec:results_det_baseline}

Before introducing market price uncertainty, we evaluate a deterministic scenario with $\Gamma=0$ to establish a baseline for comparison. The $P_{\text{base\_base}}$ trajectory is used as the nominal market price forecast, and no explicit protection against adverse price deviations is included in the baseline.

\begin{figure}[hbtp]
  \centering
  \includegraphics[width=1.0\textwidth]{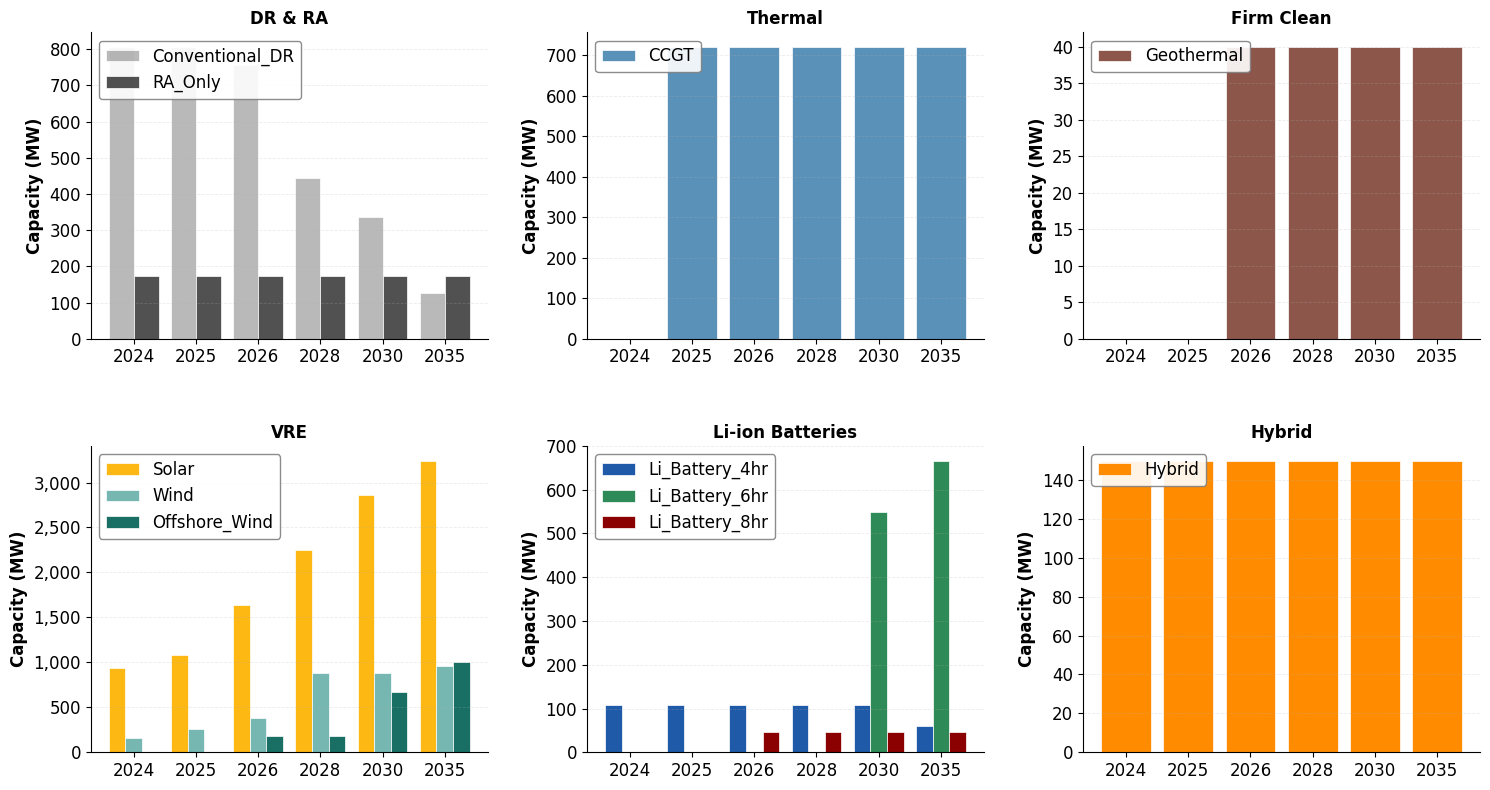}
  \caption{Capacity mix by resource category under the deterministic baseline scenario ($\Gamma=0$) across the planning horizon. The 2024 values represent existing capacity, while subsequent years report end-of-stage capacity after investment decisions and exogenous availability adjustments.}
  \label{fig:baseline_capacity_growth}
\end{figure}

\autoref{fig:baseline_capacity_growth} summarizes the end-of-stage capacity mix by resource category over the planning horizon. Capacity reported for 2024 represents the existing portfolio inherited at the beginning of the planning horizon, while subsequent years show the total available capacity, including existing resources and new investments, at the end of each planning period. Under the deterministic baseline, the portfolio satisfies all modeled CPUC IRP requirements, including the PRM, ESR, and emissions cap in each planning stage.

The transition toward a lower-carbon portfolio is most evident in the sustained expansion of variable renewable energy. Solar capacity increases from approximately 900~MW in 2024 to more than 3,200~MW by 2035, while onshore and offshore wind also expand substantially. These additions are primarily driven by the increasing ESR and declining emissions cap. Storage deployment evolves alongside renewable growth, providing flexibility to shift energy from periods of abundant renewable generation to periods of higher demand. Four-hour lithium-ion batteries provide most of the short-duration storage capacity in the early stages. Eight-hour batteries enter in 2026, while six-hour batteries are introduced in 2030 and become the dominant storage technology by 2035 as the need for longer-duration energy shifting increases.

The decline in demand response and four-hour battery capacity in later stages does not represent endogenous retirement. Instead, it results from exogenous availability assumptions that limit the continued availability of these resources. Thermal capacity is added in the first planning period and remains unchanged throughout the horizon. Geothermal capacity enters in 2026 and remains constant thereafter, providing firm clean capacity that supports system adequacy and compliance with emissions constraints. Overall, the deterministic solution follows the clean-energy pathway defined by the IRP requirements, but its procurement decisions rely entirely on the assumed nominal market price trajectory. 

\begin{figure}[htbp]
  \centering
  \subfloat[]{
    \includegraphics[width=0.6\textwidth]{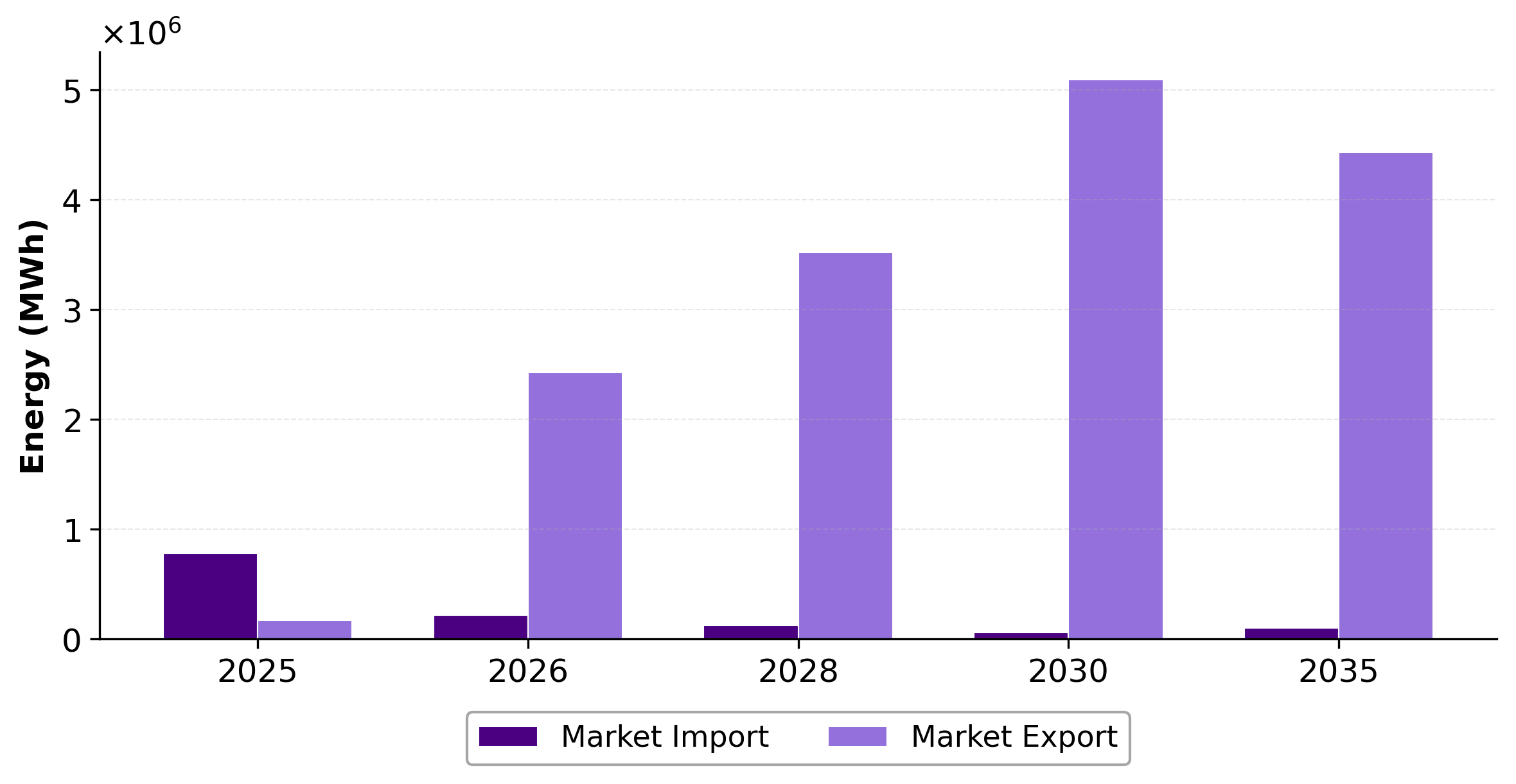}
    \label{fig:det_baseline_market_transaction}
  }
  \hfill
  \subfloat[]{
    \includegraphics[width=0.6\textwidth]{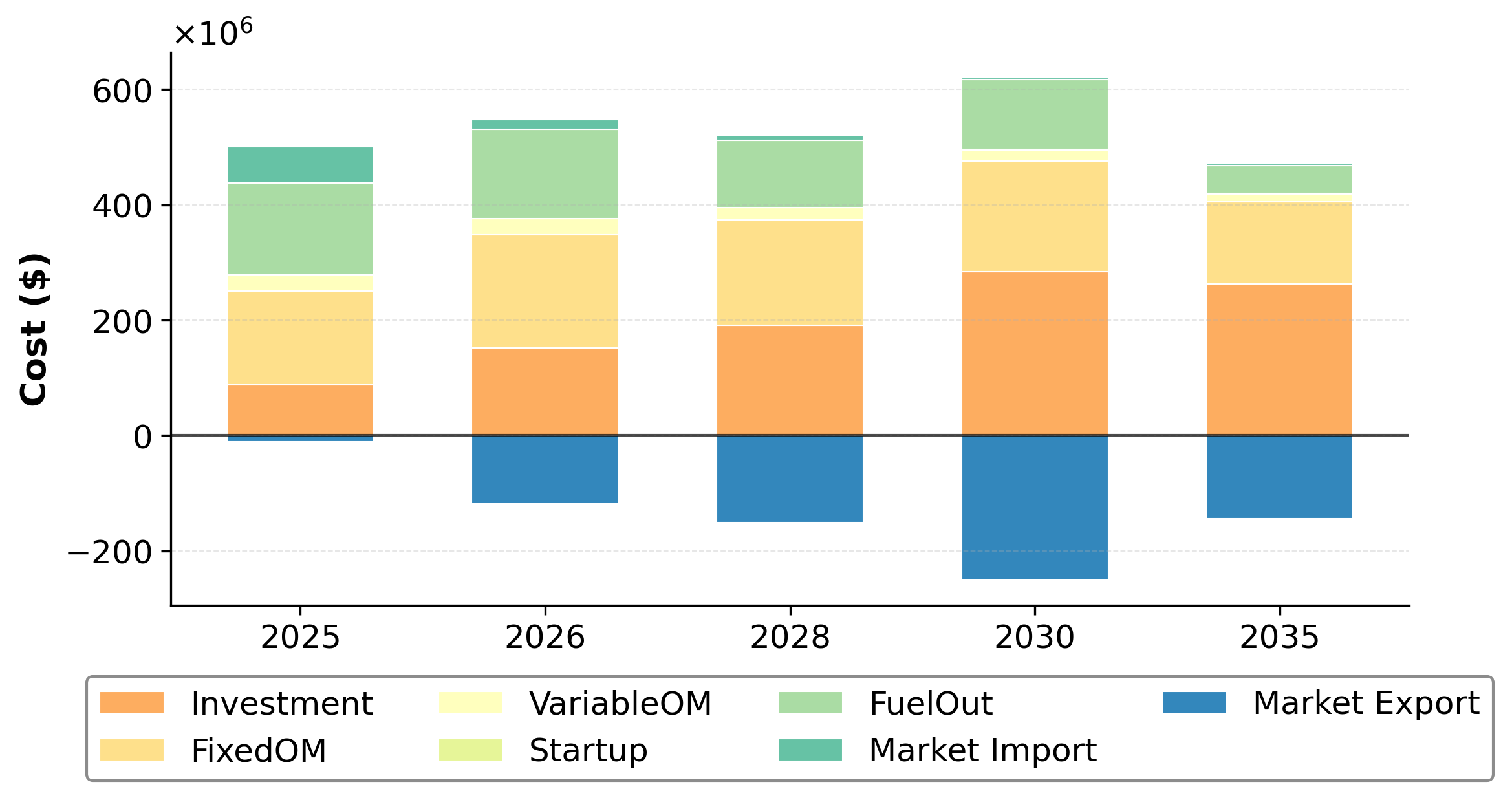}
    \label{fig:det_baseline_cost_breakdown}
  }
  \caption{Deterministic baseline results by planning year. 
  (a) Annual market import and export volumes. 
  (b) First-year cost breakdown by component, with market export revenues shown as negative values.}
   \label{fig:det_baseline_cost_and_market}
\end{figure}

These capacity additions lead to a substantial shift in market transactions, as shown in \autoref{fig:det_baseline_market_transaction}. Market imports decline sharply after 2025, while exports increase substantially as renewable capacity expands. From 2026 onward, annual exports exceed imports, indicating a shift from net importing to net exporting under the nominal price forecast. Export volumes reach their highest level in 2030 before declining slightly in 2035, while remaining well above import volumes. This pattern is consistent with increasing low-marginal-cost renewable generation, which creates recurring periods of surplus electricity available for export.

\autoref{fig:det_baseline_cost_breakdown} presents the discounted cost components for the first year of each planning period, allowing consistent comparison across periods of different lengths. The cost composition changes as the portfolio expands, with investment and fixed O\&M costs accounting for most positive system costs. Investment expenditures reach their highest level in 2030, corresponding to substantial additions of renewable generation and longer-duration storage, before declining in 2035.

Market export revenues, shown as negative costs, increasingly offset investment and operating expenditures as export volumes grow. Fuel, variable O\&M, and market import costs generally decline as renewable generation supplies a larger share of demand and reliance on thermal generation and market purchases decreases. Consequently, the deterministic portfolio partially finances its continued expansion through export revenues valued at the nominal price forecast, creating an important source of exposure when future market prices are uncertain.

\subsection{Sensitivity of Portfolio Composition to Market Price Uncertainty}

\subsubsection{Portfolio Composition}
In this subsection, we examine how portfolio composition responds to increasing market price uncertainty by varying the numerical uncertainty budget $\Gamma$, reported as a percentage of the total number of uncertain price parameters, from 0\% (deterministic) to 50\% (the most conservative effective setting). The analysis is capped at 50\% because imports and exports do not occur simultaneously in the evaluated solutions, meaning that only one of the two corresponding price deviations can affect realized market transactions in a given time step. Once the uncertainty budget covers all active market exposure, additional budget does not change the worst-case realization faced by the system. Portfolio decisions therefore converge, and further increases in $\Gamma$ produce no additional structural changes.

\begin{figure}[htbp]
  \centering
  \includegraphics[width=0.95\textwidth]{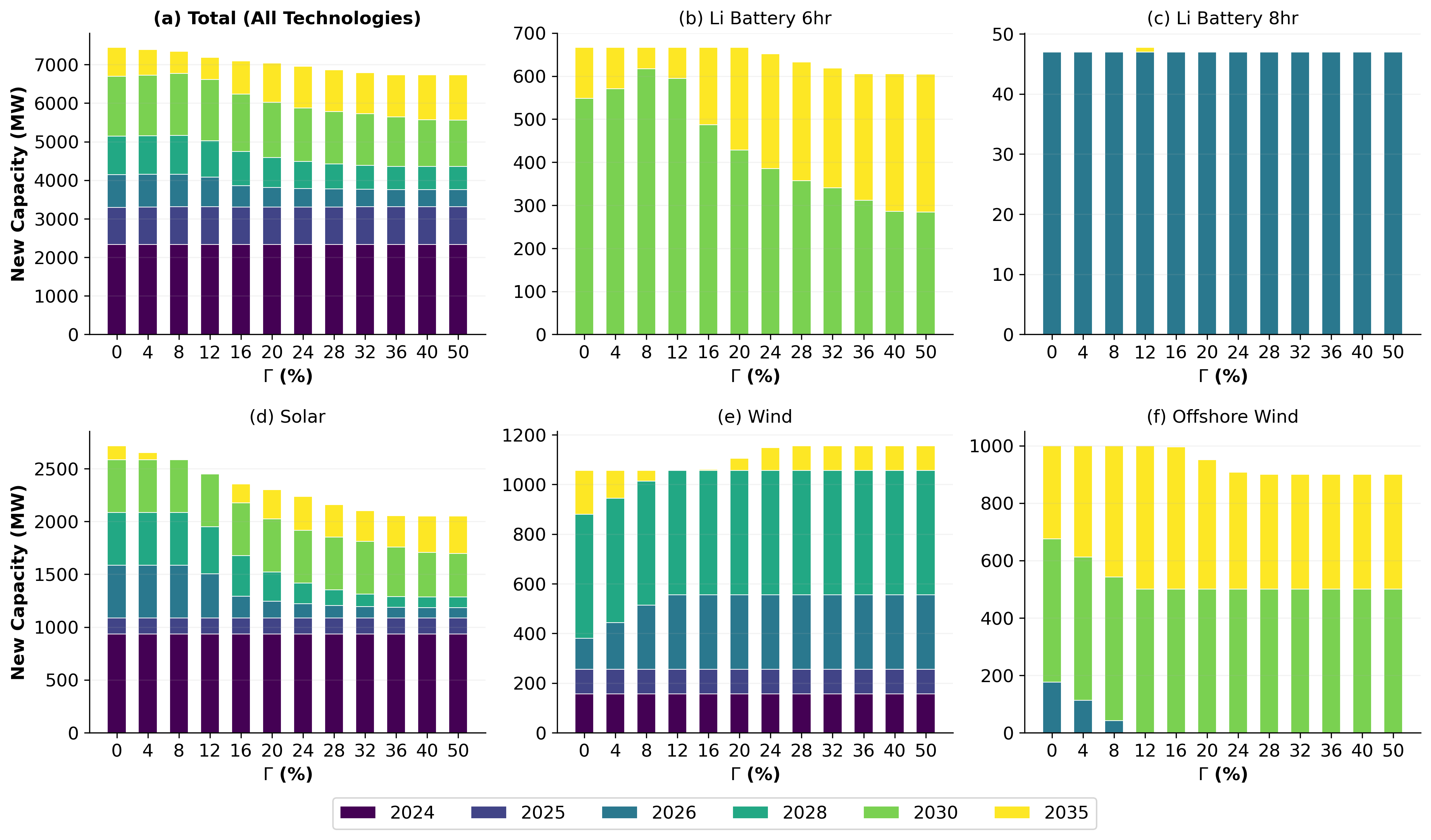}
  \caption{
    New-build capacity across the planning horizon as a function of the uncertainty budget $\Gamma$. 
    Stacked bars show capacity additions by planning year for (a) all technologies, 
    (b) 6-hour Li-ion battery, (c) 8-hour Li-ion battery, (d) solar, 
    (e) wind, and (f) offshore wind.
    }
\label{fig:baseline_capacity_part_by_gamma}
\end{figure}

\autoref{fig:baseline_capacity_part_by_gamma} summarizes total capacity procurement as a function of the uncertainty budget $\Gamma$ for the technologies most responsive to market price uncertainty. Overall procurement declines as protection against uncertainty increases. This reduction reflects weaker incentives to build generation capacity whose economic value depends on export revenues, which become less reliable at higher uncertainty budgets. Technologies exhibiting minimal sensitivity to $\Gamma$ are omitted from the main discussion; complete technology-level capacity results by planning year are provided in Appendix~\ref{app:sm}.

At the technology level, the strongest responses are observed in solar and 6-hour battery capacity. Solar additions decline steadily as $\Gamma$ increases, while onshore wind capacity increases modestly and offshore wind capacity declines slightly. Six-hour battery capacity also decreases overall as renewable buildout declines, with a larger share of its additions deferred to later planning years. Four- and 8-hour battery capacities remain comparatively stable across uncertainty budgets. Because all portfolios continue to satisfy the ESR and emissions constraints, the reduction in renewable procurement does not represent a weakening of the CCA's decarbonization requirements. Instead, it reflects a reduced incentive to build capacity beyond those requirements when its economic value depends on uncertain export revenues.

\begin{figure}[htbp]
  \centering
  \includegraphics[width=0.95\textwidth]{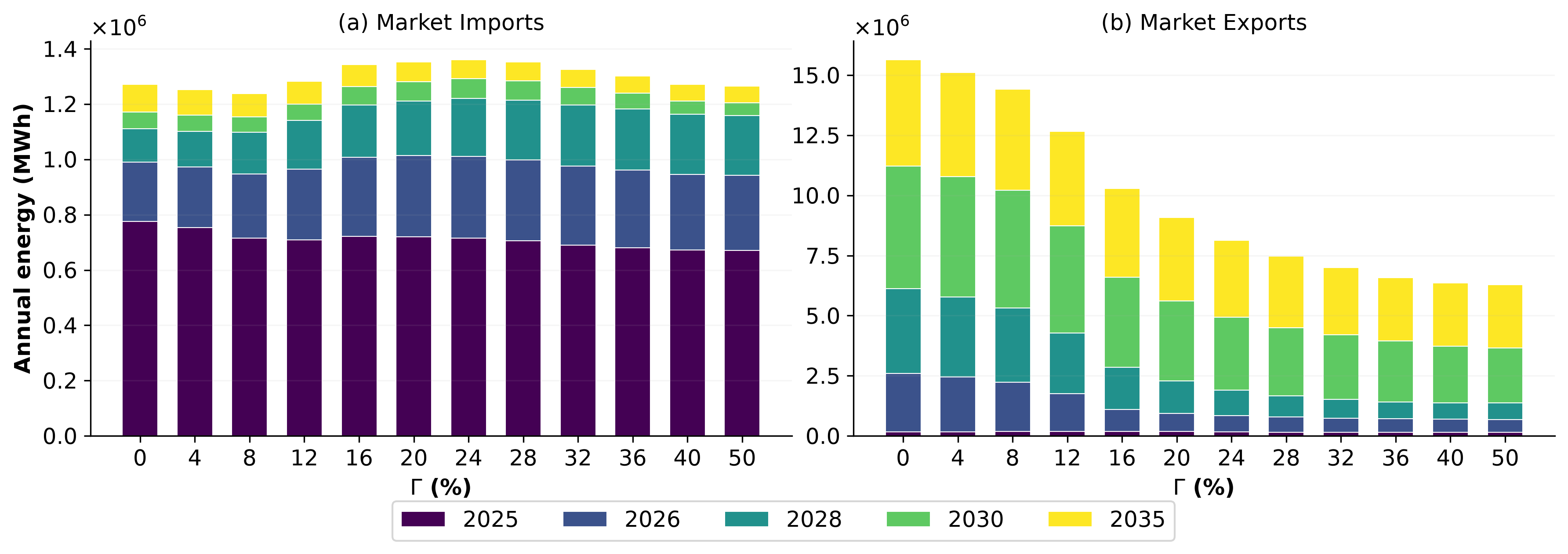}
  \caption{Annual market transactions as a function of the uncertainty budget $\Gamma$: (a) market imports; (b) market exports, reported by planning year.}
  \label{fig:baseline_gamma_markets}
\end{figure}

Imports remain relatively stable across uncertainty budgets because they primarily support demand satisfaction and system adequacy, whereas exports represent discretionary surplus energy whose economic value depends on uncertain future market prices. As shown in \autoref{fig:baseline_gamma_markets}, export volumes decline sharply with increasing $\Gamma$ across all planning years, while import volumes exhibit comparatively limited variation. This asymmetric response indicates that increasing price uncertainty primarily reduces the value of market participation as an outlet for surplus generation rather than substantially changing reliance on imports. Consequently, the incentive to invest in export-oriented capacity declines, reinforcing the reduction in total procurement observed in \autoref{fig:baseline_capacity_part_by_gamma}.

\subsubsection{Cost Trade-offs and Price of Robustness}
\begin{figure}[htbp]
  \centering
  \includegraphics[width=1\textwidth]{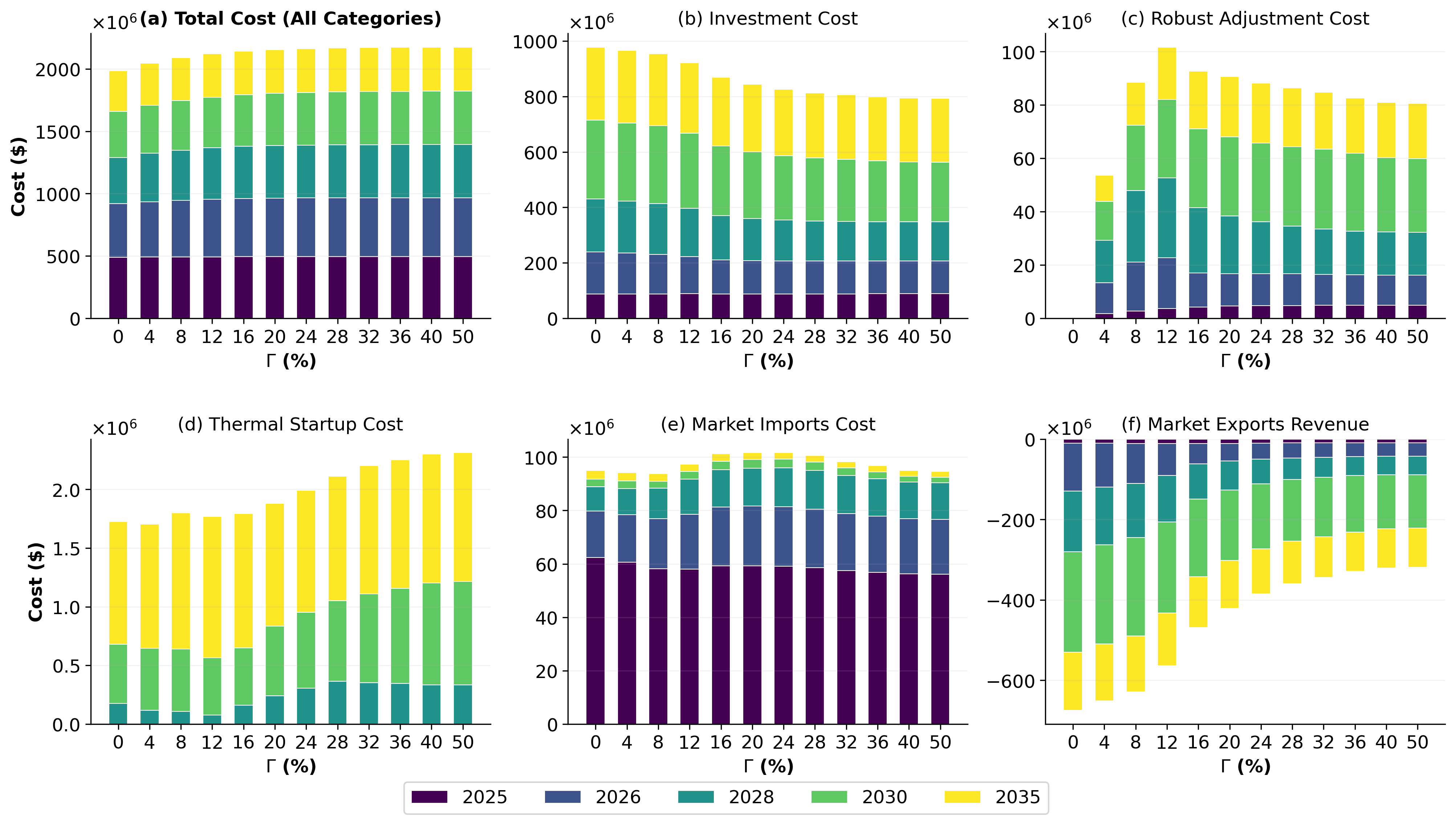}
  \caption{Evolution of major cost components with increasing uncertainty budget $\Gamma$, evaluated for the first year of each planning period.}
  \label{fig:baseline_by_gamma_cost}
\end{figure}

\autoref{fig:baseline_by_gamma_cost} illustrates how the main cost components change as the uncertainty budget $\Gamma$ increases. Total system cost rises smoothly before stabilizing at higher uncertainty budgets, indicating that additional protection becomes progressively less costly at the margin. The increase is driven primarily by the decline in market export revenues as robust portfolios reduce surplus generation and become less dependent on uncertain market sales. Market import costs remain comparatively stable, consistent with the limited change in import volumes discussed previously.

Part of the lost export revenue is offset by lower investment expenditures because increasing $\Gamma$ discourages export-oriented capacity expansion. The Robust Adjustment Cost represents the worst-case market price adjustment added to the objective by the robust formulation. This term initially increases as more adverse price deviations are allowed, reaches its highest level at an intermediate uncertainty budget, and then declines as the portfolio reduces its exposure to market transactions. Other operating cost components, including thermal startup costs, are small relative to investment costs and market revenues and do not materially alter the overall trade-off. Thus, the cost of robustness is primarily associated with forgone export opportunities and explicit protection against adverse prices, partly offset by lower capacity investment. While this cost decomposition identifies the sources of the change in system cost, it does not directly measure the total premium associated with adopting a more conservative procurement strategy. We quantify this premium using the Price of Robustness below.

Unlike the Robust Adjustment Cost, which captures the explicit worst-case price adjustment included in the objective, the Price of Robustness (PoR) measures the total economic premium associated with the investment and operational decisions adopted at a given uncertainty budget. For the baseline strategy, in which capacity is re-optimized at each value of $\Gamma$, PoR is defined as

\begin{equation}
\text{PoR}(\Gamma) \;=\; \frac{Z(\Gamma) - Z(0)}{Z(0)},
\end{equation}
where $Z(\Gamma)$ is the optimal net present system cost at uncertainty budget $\Gamma$, and $Z(0)$ is the corresponding deterministic cost.

To examine how capacity decisions influence this cost premium, we also conduct a set of \emph{fixed-capacity} experiments. For each reference budget $\Gamma^{\text{cap}}\in\{0\%,8\%,20\%,40\%\}$, capacity is first optimized at $\Gamma=\Gamma^{\text{cap}}$ and then held fixed while the operational uncertainty budget varies over the full range of $\Gamma$. For each fixed-capacity portfolio, PoR is calculated relative to that portfolio's own cost at $\Gamma=0$:

\begin{equation}
\text{PoR}_{\Gamma^{\text{cap}}}(\Gamma)
=
\frac{
Z_{\Gamma^{\text{cap}}}(\Gamma)
-
Z_{\Gamma^{\text{cap}}}(0)
}{
Z_{\Gamma^{\text{cap}}}(0)
}.
\label{eq:por_fixed_capacity}
\end{equation}

\begin{figure}[htbp]
  \centering
  \subfloat[]{
    \includegraphics[width=0.48\textwidth]{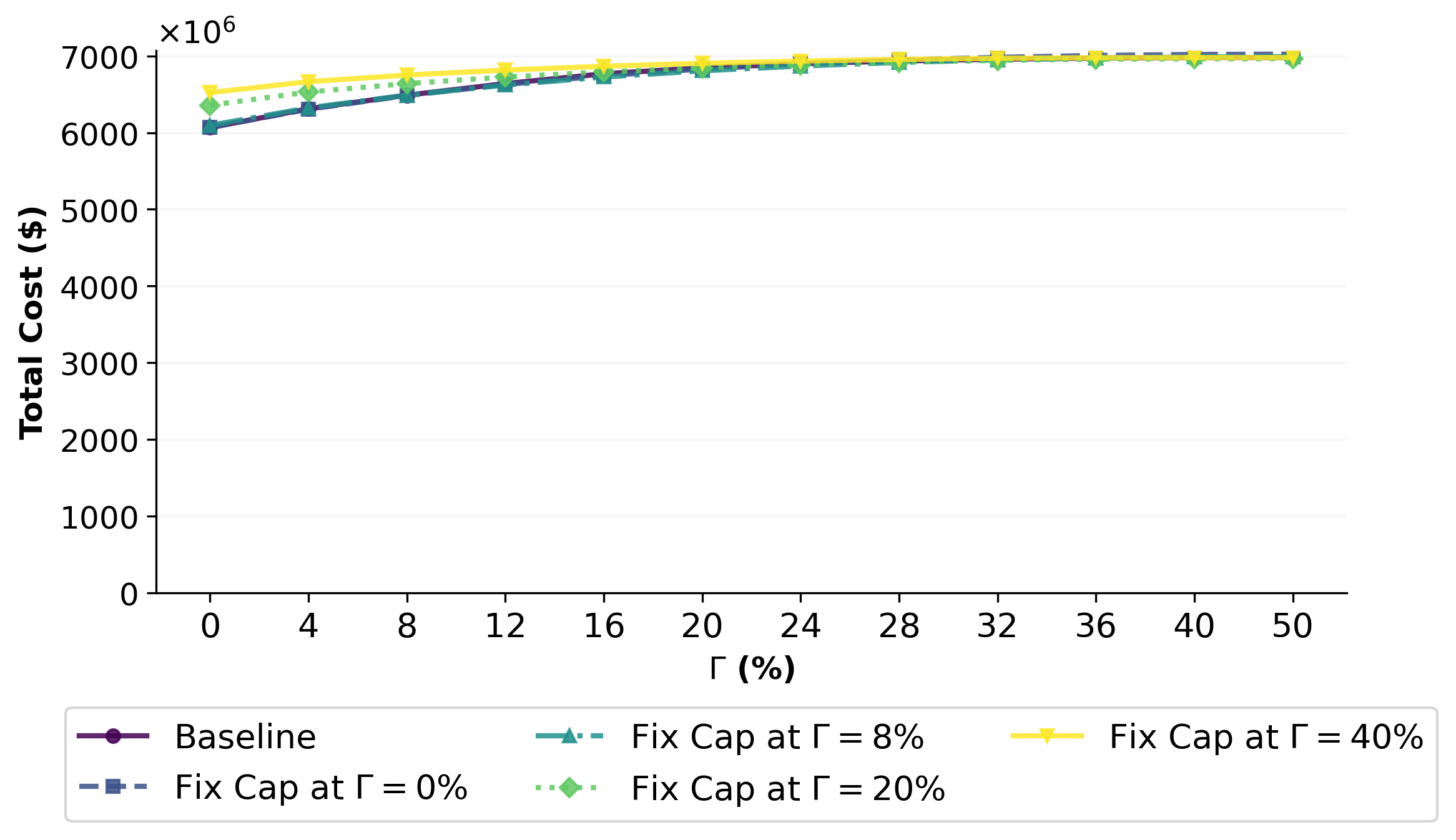}
    \label{fig:por_totalcost}
  }\hfill
  \subfloat[]{
    \includegraphics[width=0.48\textwidth]{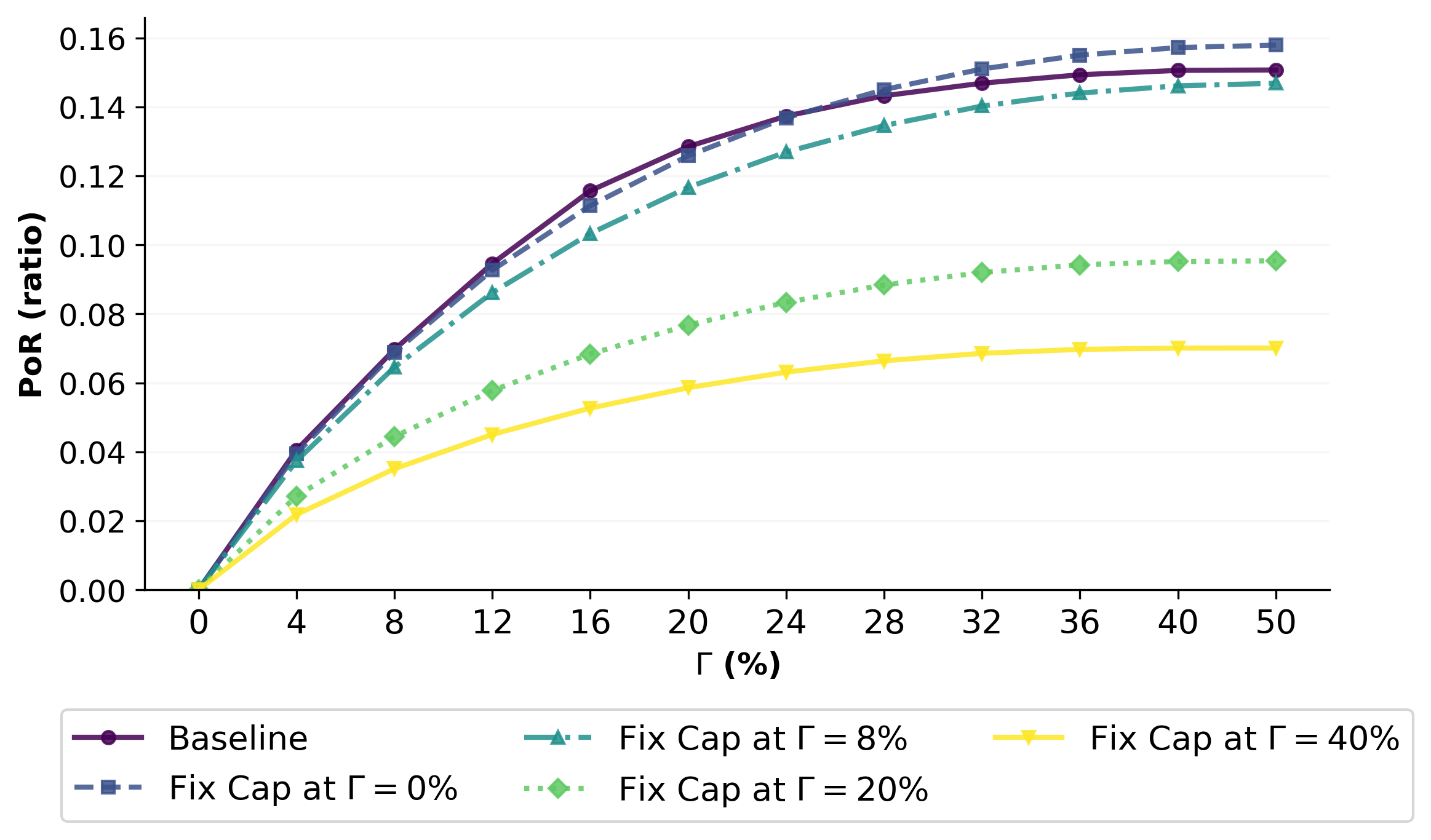}
    \label{fig:por_ratio}
  }
  \caption{
  Robustness–cost trade-offs under hourly market price uncertainty. 
  (a) Net present system cost as a function of the uncertainty budget $\Gamma$ for the baseline case (capacity re-optimized for each $\Gamma$) and for scenarios with capacity fixed at portfolios selected at $\Gamma=0\%, 8\%, 20\%$ and $40\%$. 
  (b) Corresponding Price of Robustness (PoR)
  }
  \label{fig:por_totalcost_por}
\end{figure}

The resulting total system costs and PoR values are shown in \autoref{fig:por_totalcost_por}.

For the baseline strategy, PoR increases rapidly at low and moderate uncertainty budgets, reaching approximately 14\% by $\Gamma=28\%$ and about 15\% at the highest budgets. The curve gradually flattens, indicating that the marginal cost of additional protection declines as active market exposure becomes increasingly covered.

The fixed-capacity experiments show that total system costs become increasingly similar as the operational uncertainty budget rises. Portfolios selected at higher values of $\Gamma^{\text{cap}}$ begin with higher nominal costs but exhibit smaller relative cost increases as operational uncertainty grows. Capacity fixed at $\Gamma^{\text{cap}}=0\%$ shows the greatest sensitivity, whereas capacity fixed at $\Gamma^{\text{cap}}=8\%$ provides a modest reduction in PoR with little change in nominal cost. Portfolios fixed at $\Gamma^{\text{cap}}=20\%$ and $40\%$ are less sensitive to increasing operational uncertainty, but this lower relative sensitivity is achieved through higher upfront costs.

These results demonstrate the trade-off between nominal affordability and protection against adverse market prices. Moderate hedging can reduce cost sensitivity without substantially increasing the initial portfolio cost, while more conservative capacity decisions provide additional protection at a higher nominal premium. The \autoref{ssec:out_of_sample} out-of-sample analysis presented below evaluates whether these additional costs translate into improved performance under alternative market price realizations.

\subsubsection{Operational Implications of Robust Capacity Planning}
\label{sssec:operation_by_gamma}

\begin{figure}[htbp]
  \centering
   \subfloat[]{
    \includegraphics[width=0.6\linewidth]{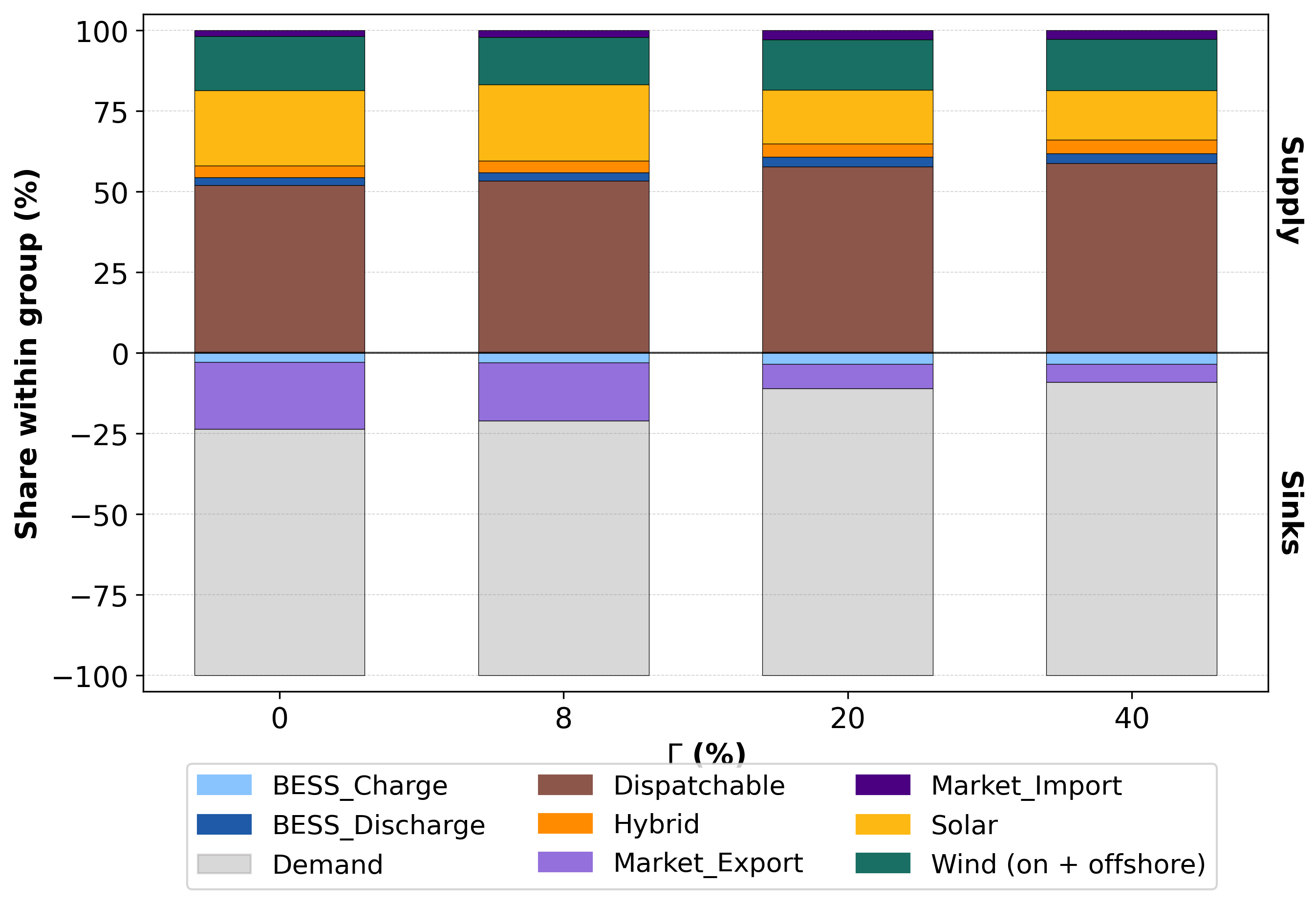}
    \label{fig:baseline_dispatch_shares26}
  }\hfill
  \subfloat[]{
    \includegraphics[width=0.6\textwidth]{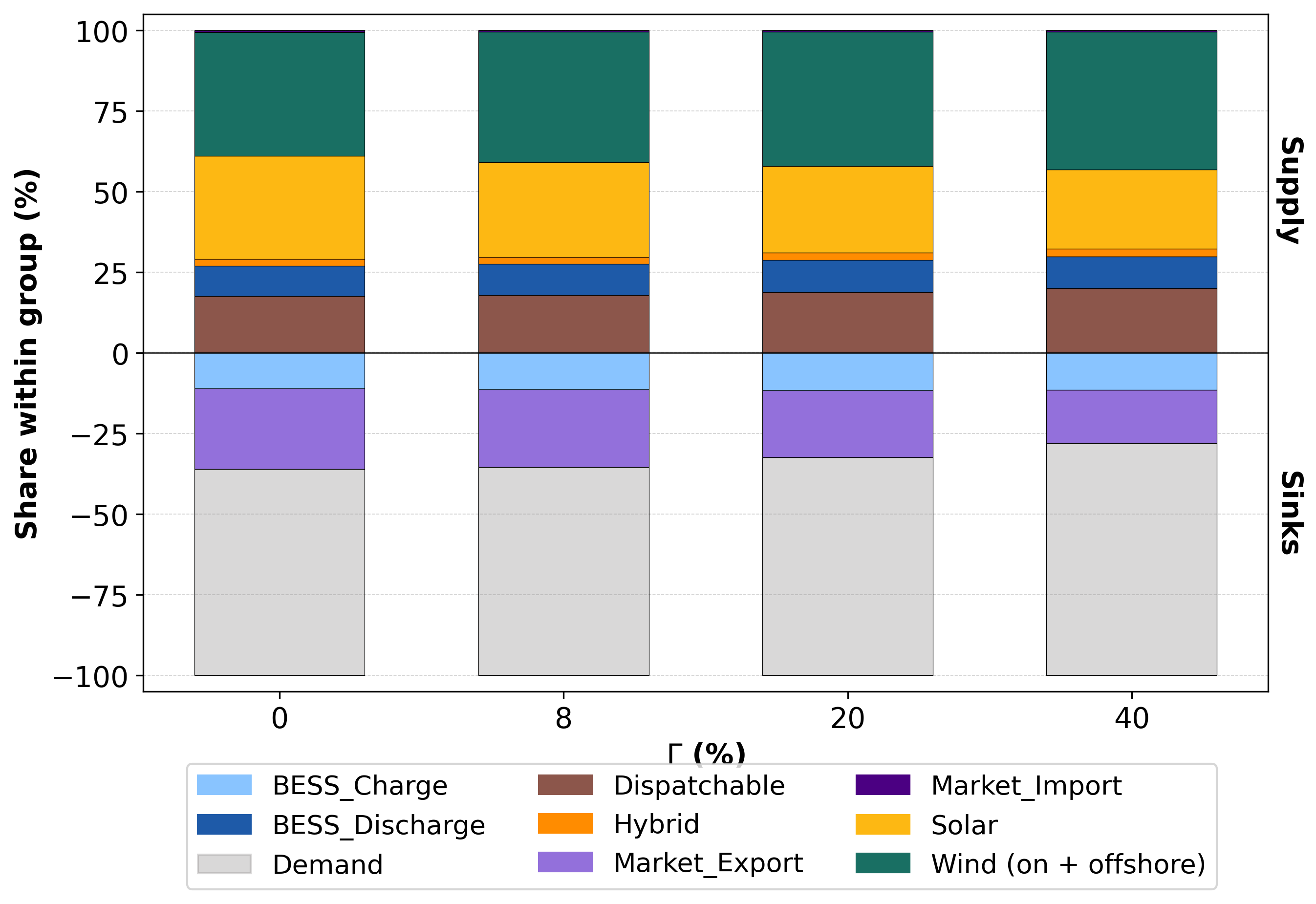}
    \label{fig:baseline_dispatch_shares35}
  }
  \caption{
  Operational dispatch shares under increasing market price uncertainty for selected uncertainty budgets $\Gamma \in \{0,8,20,40\}$. 
  Bars are normalized within each $\Gamma$, with positive values indicating supply and negative values indicating sinks. 
  Dispatchable generation aggregates CCGT and geothermal resources, storage aggregates 4-, 6-, and 8-hour batteries, and wind aggregates onshore and offshore resources. 
  (a) 2026 (early-stage portfolio).
  (b) 2035 (end-state portfolio).
  }  
  \label{fig:baseline_dispatch_shares}
\end{figure}

While investment decisions determine the available resource mix, increasing market price uncertainty also affects how installed capacity is utilized operationally. To examine these effects, we analyze aggregated dispatch shares for selected uncertainty budgets $\Gamma \in \{0,8,20,40\}$, normalized within each $\Gamma$ such that positive values indicate supply contributions and negative values indicate sinks (\autoref{fig:baseline_dispatch_shares}).

As the uncertainty budget increases, the robust solution shifts operational reliance toward resources that are not directly exposed to market prices, particularly wind generation. This behavior is evident in both the early-stage portfolio in 2026 (\autoref{fig:baseline_dispatch_shares26}) and the later-stage portfolio in 2035 (\autoref{fig:baseline_dispatch_shares35}). Total wind capacity changes comparatively little across uncertainty budgets, suggesting that the observed shift is influenced largely by how available wind capacity is utilized. Reliance on market exports also decreases as $\Gamma$ increases, consistent with the reduction in absolute export volumes reported in \autoref{fig:baseline_gamma_markets}. Instead, the system relies more heavily on internally available firm and renewable resources to meet demand and manage variability.

\begin{figure}[htbp]
  \centering
  \includegraphics[width=0.75\linewidth]{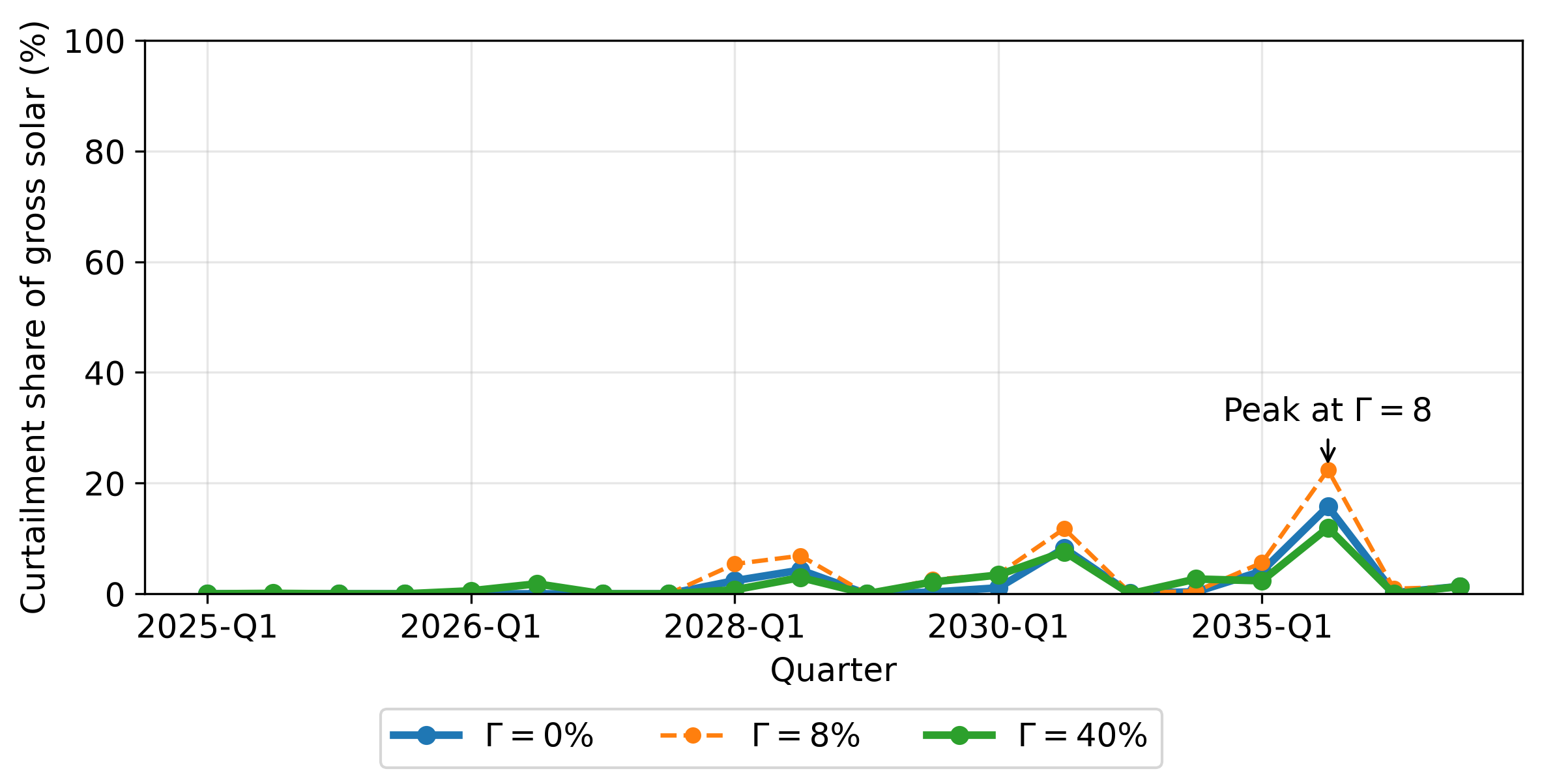}
  \caption{Quarterly share of gross solar availability curtailed during midday hours from 10:00 to 15:00 under selected uncertainty budgets $\Gamma\in\{0\%,8\%,40\%\}$.}
  \label{fig:midday_export_curtailment}
\end{figure}

The reduced reliance on market exports raises the question of how surplus renewable energy, particularly solar generation during peak production hours, is managed as robustness increases. To examine this effect, we analyze solar curtailment during midday hours from 10:00 to 15:00. \autoref{fig:midday_export_curtailment} shows the percentage of gross solar availability curtailed, aggregated quarterly for selected uncertainty budgets. For visual clarity, results are shown for $\Gamma\in\{0\%,8\%,40\%\}$, as $\Gamma=20\%$ exhibits intermediate behavior.

As uncertainty aversion increases, reduced export activity initially coincides with higher solar curtailment. At the intermediate uncertainty budget of $\Gamma=8\%$, solar capacity remains substantial while the use of exports to absorb surplus generation has already declined. At $\Gamma=40\%$, curtailment decreases as the robust portfolio reduces solar capacity and therefore limits exposure to surplus generation. This non-monotonic pattern shows how robustness affects both renewable capacity decisions and the operational management of surplus solar generation.

\subsection{Sensitivity of Procurement Decisions to Uncertainty Set Specification}

To assess the sensitivity of robust procurement decisions to the specification of the market price uncertainty set, we compare solutions obtained under alternative constructions for unspecified import prices. In the baseline specification (\texttt{base\_base}), the nominal price trajectory is given by $P_{\text{base\_base}}$, as described in \autoref{sec:case_study}, with uncertainty bounds defined by asymmetric deviations around this profile that preserve the temporal structure of the underlying forecast.

In the alternative specification (\texttt{base\_flat}), the nominal trajectory is instead given by $P_{\text{base\_flat}}$. Uncertainty bounds are constructed using the minimum and maximum values observed across all six vendor projections at each hour, with deviations from the nominal price constrained to be at least \$5/MWh. This approach produces a higher nominal price level and wider uncertainty envelopes in later planning years. Robust procurement experiments are conducted across increasing values of the uncertainty budget $\Gamma$ under both specifications to isolate the effect of uncertainty-set design on investment outcomes.

Although both uncertainty sets are derived from the same ensemble of vendor-provided trajectories, changing the nominal price profile alters the temporal alignment of adverse price deviations relative to system operations. To further assess sensitivity to potential underestimation of price uncertainty, we also evaluate a stressed-envelope variant in which the \texttt{base\_flat} deviation bounds are uniformly scaled by a factor $\kappa=1.5$. This stress test is not intended to represent a specific forecast scenario, but rather to examine the sensitivity of procurement outcomes to systematically larger price deviations.

The analysis focuses on end-of-horizon installed capacity, which reflects the cumulative procurement decisions made over the planning horizon.
 
 \begin{figure}[htbp]
  \centering
  \includegraphics[width=1\textwidth]{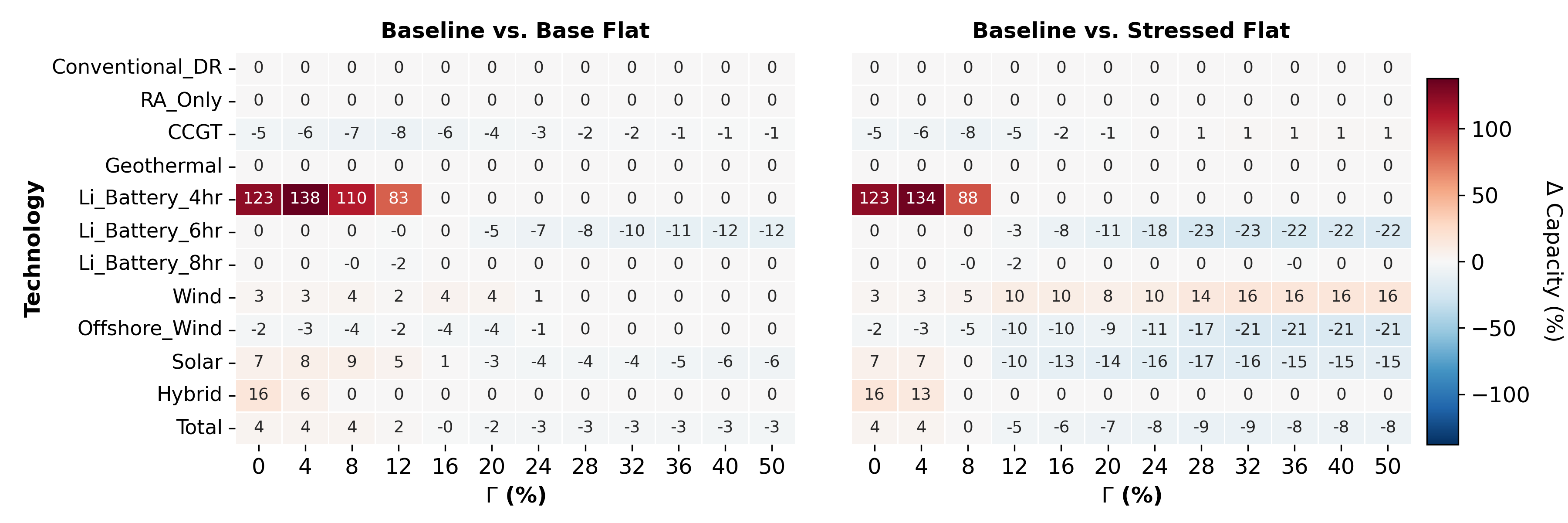}  
    \caption{Sensitivity of end-of-horizon installed capacity to uncertainty-set specification. Heatmaps report percentage differences by technology relative to the \texttt{base\_base} solution at each uncertainty budget $\Gamma$. The left panel compares the \texttt{base\_flat} specification, while the right panel applies the stressed deviation envelope (\texttt{stressed\_flat}, $\kappa=1.5$).}
  \label{fig:baseline_mean_price_delta_end_capacity}
\end{figure}
 
\Autoref{fig:baseline_mean_price_delta_end_capacity} compares end-of-horizon installed capacity under the alternative uncertainty-set specifications, with differences reported relative to the \texttt{base\_base} solution at each value of $\Gamma$. Under the \texttt{base\_flat} specification, the largest differences occur at low uncertainty budgets and generally decrease as $\Gamma$ increases. The \texttt{stressed\_flat} specification produces more persistent differences across the full range of uncertainty budgets, indicating greater sensitivity to the width of the uncertainty envelope.

At low values of $\Gamma$, 4-hour lithium-ion battery capacity exhibits the largest percentage differences, reflecting the sensitivity of this comparatively small capacity category to changes in the nominal price trajectory. Under \texttt{base\_flat}, this difference disappears at moderate uncertainty budgets, while 6-hour battery and solar capacities decline modestly at higher values of $\Gamma$. The \texttt{stressed\_flat} specification produces larger and more persistent changes, including reductions in 6-hour battery, offshore wind, and solar capacity, accompanied by increased onshore wind capacity. These results indicate that wider deviation bounds affect both storage selection and the composition of renewable procurement.

Despite these technology-level changes, differences in total installed capacity remain comparatively moderate. Relative to \texttt{base\_base}, total capacity under both alternative specifications is approximately 4\% higher at low uncertainty budgets. At higher budgets, total capacity is approximately 3\% lower under \texttt{base\_flat} and 8\% lower under \texttt{stressed\_flat}. Thus, changes in uncertainty-set specification affect the allocation of capacity across technologies more strongly than the overall scale of procurement.

\begin{figure}[htbp]
  \centering
  \includegraphics[width=0.95\textwidth]{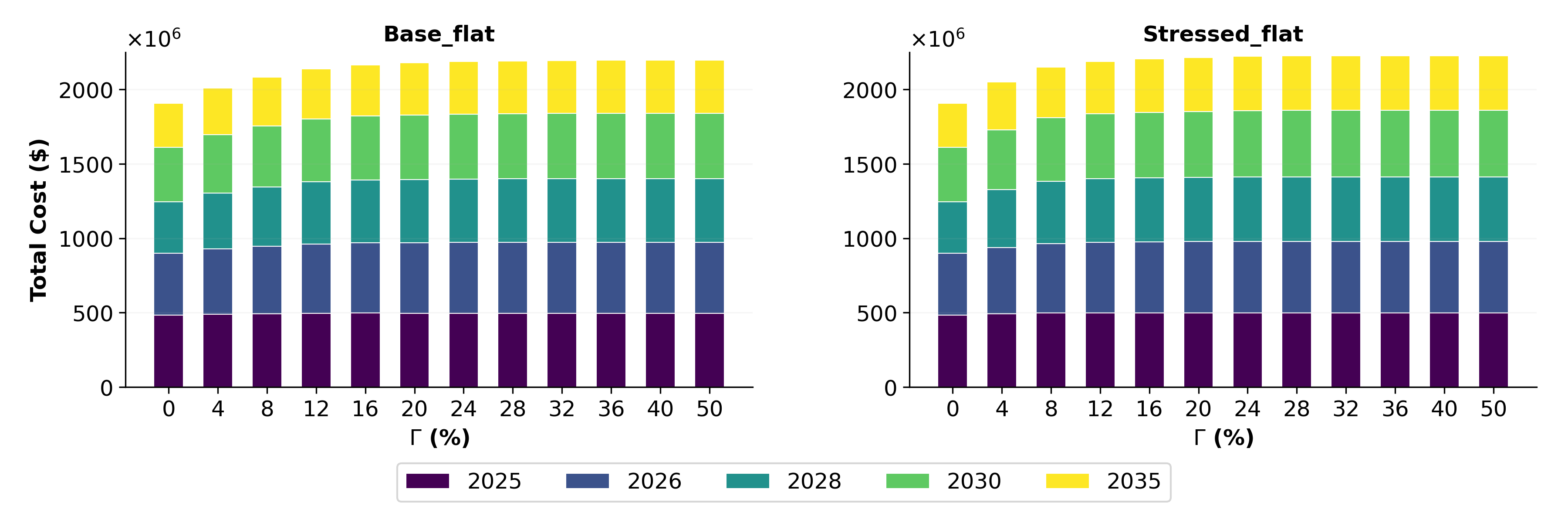}  
    \caption{Total system cost by planning period as a function of the uncertainty budget $\Gamma$ under the \texttt{base\_flat} and \texttt{stressed\_flat} uncertainty-set specifications.}
  \label{fig:baseline_uncertainty_set_cost}
\end{figure}

\autoref{fig:baseline_uncertainty_set_cost} examines whether these differences in uncertainty-set construction translate into substantial economic effects. Under both \texttt{base\_flat} and \texttt{stressed\_flat}, total system cost increases smoothly with $\Gamma$ before stabilizing at moderate and high uncertainty budgets. The \texttt{stressed\_flat} specification produces slightly higher costs because of its wider deviation envelope, but the cost difference remains relatively small compared with the observed technology-level capacity changes. This indicates that alternative uncertainty-set specifications can produce different procurement portfolios within a comparatively narrow range of total system cost.

Overall, procurement decisions are moderately sensitive to the choice of nominal price trajectory and more persistently sensitive to the width of the uncertainty envelope. These specifications can materially change the allocation of capacity across renewable and storage technologies while producing comparatively small differences in total system cost. This highlights the importance of testing alternative uncertainty-set specifications when evaluating long-term CCA procurement strategies.

\subsection{Out-of-Sample Robustness Validation}
\label{ssec:out_of_sample}
To evaluate the robustness of CCA procurement decisions under unseen market conditions, we perform an out-of-sample Monte Carlo validation using 150 synthetic price trajectories. Each trajectory is initialized as a random convex combination of the six vendor-provided price projections, generating a diverse set of baseline futures. A persistent AR(1) deviation process is then applied to preserve realistic temporal autocorrelation, allowing smooth multi-hour price movements while avoiding implausible hour-to-hour volatility. All simulated trajectories are constrained within a soft envelope derived from the original projections to maintain plausible price ranges. In addition, a scarcity-jump profile is constructed using a Bernoulli-triggered process to represent rare but extreme price spikes associated with grid-stress events.

In total, 157 price profiles are evaluated, comprising the 150 synthetic trajectories, the six original vendor profiles, and the scarcity-jump profile. For the deterministic portfolio ($\Gamma=0\%$), three runs did not solve within the allocated time limit. One run reached the time limit at $\Gamma=8\%$, while all runs solved successfully at $\Gamma=40\%$. Performance metrics for each uncertainty budget are calculated over the successfully solved runs, and time-limit terminations are reported separately. 

\begin{figure}[htbp]
  \centering
   \subfloat[]{
    \includegraphics[width=0.48\linewidth]{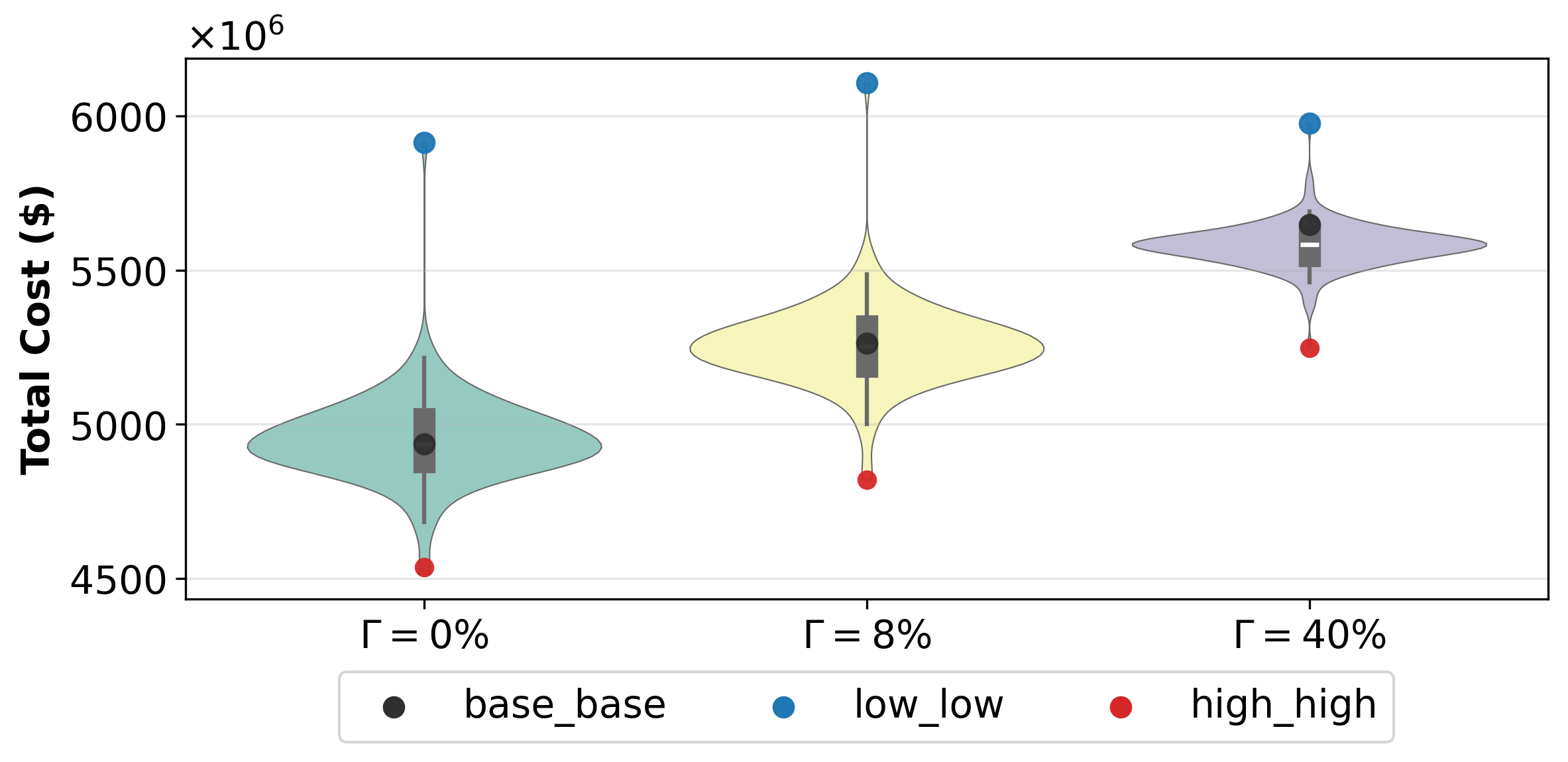}
    \label{fig:oos_total_cost_violin}
  }\hfill
  \subfloat[]{
    \includegraphics[width=0.48\textwidth]{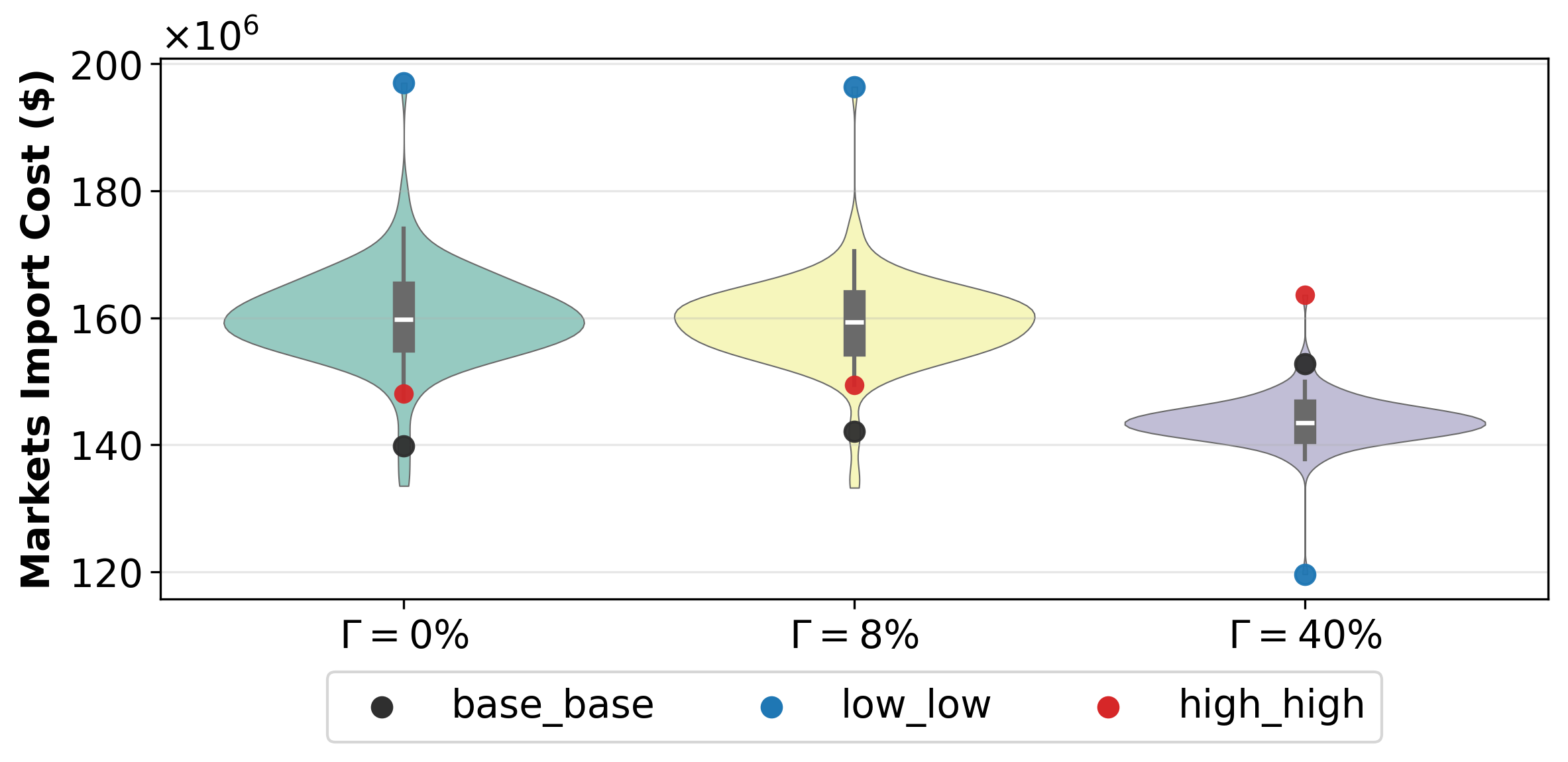}
    \label{fig:oos_import_cost_violin}
  }
 \caption{
  Out-of-sample cost distributions under unseen market price realizations.
  Panel (a) shows the distribution of total system cost, while panel (b) reports market import cost, for selected uncertainty budgets $\Gamma \in \{0,8,40\}$. 
  Violins summarize the distribution across all solved price scenarios, with markers indicating the nominal trajectory (base\_base) and the extremal profiles (low\_low and high\_high) used to construct the uncertainty set.
  }
  \label{fig:oos_violin}
\end{figure}

\autoref{fig:oos_violin} summarizes the distributions of total system cost and market import cost across the out-of-sample price realizations. \autoref{fig:oos_total_cost_violin} shows that increasing robustness shifts the total cost distribution upward while reducing its dispersion. Although absolute upper-tail costs increase with $\Gamma$, the difference between tail outcomes and the corresponding nominal cost narrows. The extreme profiles used to construct the uncertainty set, \texttt{low\_low} and \texttt{high\_high}, bound the total cost distributions as expected, with \texttt{low\_low} consistently producing the highest system costs because of reduced export revenues.

\autoref{fig:oos_import_cost_violin} shows market import costs under the same price profiles. Unlike total system cost, import cost does not follow a uniform ordering across the extreme profiles, particularly at high robustness. As $\Gamma$ increases, market exposure is reduced and residual import costs become small relative to total system cost. Consequently, \texttt{high\_high} may produce higher import costs than \texttt{low\_low}, even as overall cost exposure becomes more stable. This difference shows that component-level costs reflect specific operational exposures rather than system-wide risk, supporting the use of total cost and tail-risk measures as the primary robustness indicators.

\Autoref{tab:oos_robustness_metrics} summarizes performance under the out-of-sample price realizations. Mean, median, 95th-percentile, and $\mathrm{CVaR}_{95}$ costs increase with $\Gamma$, showing that greater robustness raises the overall cost level. However, the difference between $\mathrm{CVaR}_{95}$ and the corresponding nominal portfolio cost narrows as $\Gamma$ increases. Thus, robust procurement does not reduce absolute tail costs, but it reduces upper-tail exposure relative to the planned cost of each portfolio. The difference in the number of successfully solved runs is reported for transparency but is not used to evaluate portfolio robustness.

\begin{table}[htbp]
\centering
\footnotesize 
\caption{Out-of-sample performance metrics under unseen market price realizations. 
Results are reported for selected uncertainty budgets $\Gamma$, aggregated over all successfully solved scenarios.}
\label{tab:oos_robustness_metrics}
\begin{tabular}{lcccccccc}
\hline
$\Gamma$ 
& Scenarios 
& Solved 
& Time limit (\%) 
& Nominal Cost
& Mean Cost 
& Median Cost 
& P95 Cost 
& CVaR$_{95}$ Cost \\
\hline
0 
& 157 
& 154 
& 1.91 
& $5.2 \times 10^9$
& $4.9 \times 10^9$
& $4.9 \times 10^9$
& $5.1 \times 10^9$
& $5.3 \times 10^9$ \\

8 
& 157 
& 156 
& 0.64 
& $5.5 \times 10^9$
& $5.3 \times 10^9$
& $5.3 \times 10^9$
& $5.4 \times 10^9$
& $5.6 \times 10^9$
 \\

40 
& 157 
& 157 
& 0.00 
& $5.8 \times 10^9$
& $5.6 \times 10^9$
& $5.6 \times 10^9$
& $5.7 \times 10^9$
& $5.8 \times 10^9$
 \\
\hline
\end{tabular}
\end{table}

\subsection{Policy and Decision Implications}

A central practical question in applying robust optimization is how a planner should select the uncertainty budget $\Gamma$. Increasing $\Gamma$ provides greater protection against adverse market price realizations, but it can also alter the procurement portfolio and increase planning costs. The appropriate level of robustness therefore depends on the planner's tolerance for downside financial exposure, willingness to deviate from the baseline procurement plan, and desired reliance on external markets. To characterize these trade-offs, we evaluate three complementary dimensions: downside financial protection, portfolio stability, and market dependence. Together, these metrics provide a decision-oriented framework for comparing candidate uncertainty budgets rather than prescribing a single value of $\Gamma$.

\subsubsection{Downside Financial Protection}

For CCAs exposed to uncertain market prices, an important consideration is how much additional cost a procurement portfolio may experience under adverse price realizations relative to its nominal planning cost. We quantify this exposure using the Downside Cost Premium (DCP), defined for a portfolio planned under uncertainty budget $\Gamma^\ast$ as

\begin{equation}
\mathrm{DCP}(\Gamma^\ast)
=
\frac{
\mathrm{CVaR}_{95}(\Gamma^\ast)
-
Z_{\mathrm{nom}}(\Gamma^\ast)
}{
Z_{\mathrm{nom}}(\Gamma^\ast)
},
\label{eq:downside_cost_premium}
\end{equation}

where $Z_{\mathrm{nom}}(\Gamma^\ast)$ is the cost of the fixed portfolio under the nominal price trajectory, and $\mathrm{CVaR}_{95}(\Gamma^\ast)$ represents its average cost within the worst 5\% of successfully solved out-of-sample realizations. A positive DCP indicates that upper-tail costs exceed the nominal planning cost, whereas a negative value indicates that the nominal cost exceeds the observed $\mathrm{CVaR}_{95}$.

Based on the out-of-sample results reported in \autoref{tab:oos_robustness_metrics}, DCP declines from 1.3\% under the deterministic solution ($\Gamma^\ast=0\%$) to 0.6\% at $\Gamma^\ast=8\%$ and becomes slightly negative ($-0.8\%$) at $\Gamma^\ast=40\%$. Thus, a relatively small uncertainty budget removes a substantial portion of the downside financial exposure, while the highly robust portfolio is sufficiently conservative that its tail out-of-sample cost falls below its nominal planning cost.

To express this reduction relative to the initial exposure faced by the deterministic portfolio, we define hedge effectiveness (HE) as

\begin{equation}
\label{eq:hedge_effectiveness}
\text{HE}(\Gamma^\ast)
=
1 -
\frac{\text{DCP}(\Gamma^\ast)}{\text{DCP}(0)} .
\end{equation}

At $\Gamma=8\%$, hedge effectiveness is approximately 52\%, indicating that more than half of the deterministic portfolio's downside cost premium is eliminated. At $\Gamma=40\%$, HE exceeds 100\%, reflecting conservative over-hedging rather than an equivalent increase in economically meaningful risk reduction. From a planning perspective, the two metrics therefore distinguish between obtaining substantial protection against downside cost exposure and adopting increasingly conservative portfolios after that exposure has largely been mitigated.

\subsubsection{Portfolio Stability}

Financial protection alone does not determine whether a robustness level is attractive to a planner. A robust solution that substantially changes an existing procurement strategy may carry implementation, contracting, or planning implications even if it provides greater protection against uncertain prices. We therefore measure portfolio stability relative to the deterministic procurement plan as

\begin{equation}
\label{eq:portfolio_stability}
\text{Stability}(\Gamma)
=
1 -
\frac{
\sum_k
\left|
\text{Cap}_k(\Gamma)-\text{Cap}_k(0)
\right|
}{
\sum_k \text{Cap}_k(0)
}.
\end{equation}

Here, $\text{Cap}_k(\Gamma)$ denotes the installed capacity of technology $k$ under uncertainty budget $\Gamma$. A score of one indicates no change from the deterministic portfolio, while lower values indicate increasing structural deviation from that baseline.

The stability score remains close to one at $\Gamma=8\%$, indicating that the substantial improvement in downside financial protection at this level requires only modest adjustments to the baseline investment plan. By $\Gamma=40\%$, the score has declined and largely plateaued, suggesting that the portfolio has completed most of its structural adaptation to market price uncertainty. Portfolio stability therefore provides planners with a way to evaluate whether additional financial protection requires material changes to the underlying procurement strategy.

\subsubsection{Market Dependence}

A third consideration for CCAs is the extent to which the resulting portfolio depends on external market purchases to satisfy demand. Greater market dependence can leave the system more exposed to uncertain future market conditions even when expected procurement costs remain acceptable. For each out-of-sample price realization $s$, market dependence is defined as

\begin{equation}
\label{eq:market_dependence}
D_s(\Gamma)
=
\frac{
\sum_t \text{Imports}_{t,s}(\Gamma)
}{
\sum_t \text{Demand}_t
}.
\end{equation}

\Autoref{tab:market_dependence} reports the resulting mean and tail dependence measures for the fixed portfolios associated with the selected uncertainty budgets. Mean market dependence declines from 3.27\% at $\Gamma=0\%$ to 3.15\% at $\Gamma=8\%$ and 2.85\% at $\Gamma=40\%$. The reduction is more pronounced in the upper tail: $\text{CVaR}_{95}$ dependence falls from 4.52\% to 4.27\% and ultimately to 3.39\%, respectively. The narrowing gap between average and tail dependence indicates that increasing robustness reduces not only typical market reliance but also exposure to unusually high import requirements.

\begin{figure}[htbp]
\centering
\includegraphics[width=0.5\linewidth]{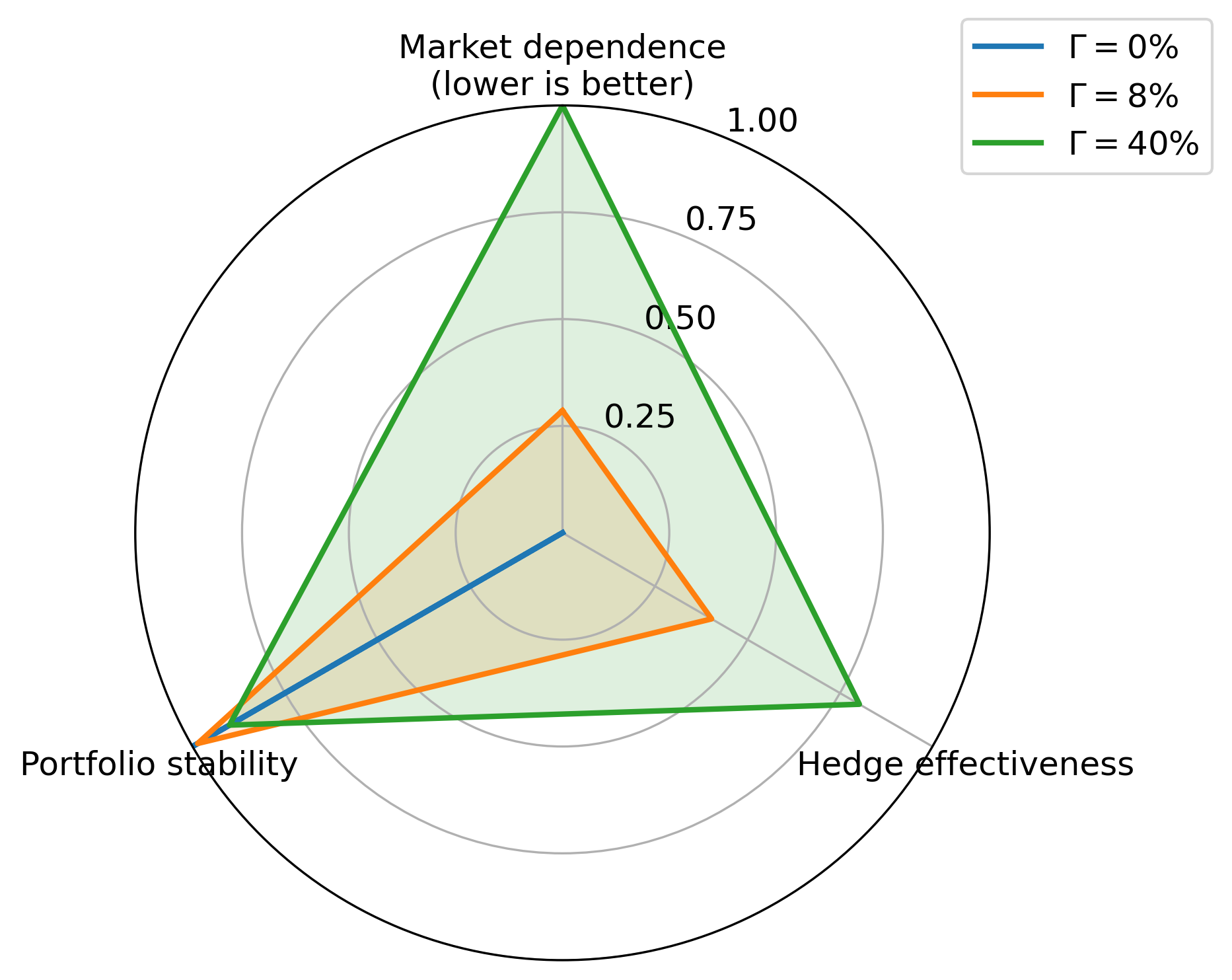}
\caption{
Decision-oriented comparison of downside financial protection, portfolio stability, and market dependence for portfolios planned at $\Gamma \in \{0,8,40\}\%$. Hedge effectiveness and market dependence are normalized for visualization, with higher values representing greater protection and lower market reliance, respectively.
}
\label{fig:decision_metrics}
\end{figure}

The combined trade-offs are summarized in \autoref{fig:decision_metrics}. Rather than identifying a universally optimal uncertainty budget, the three metrics allow planners to compare robustness levels according to their institutional priorities. A planner seeking substantial protection against adverse prices while limiting changes to the baseline procurement strategy may favor a moderate uncertainty budget, such as $\Gamma=8\%$, which eliminates approximately half of the baseline downside premium while maintaining high portfolio stability. Higher uncertainty budgets provide additional reductions in market dependence and residual tail exposure, but these improvements occur after much of the initial financial protection has already been achieved.

These trade-offs also have direct implications for the CCA's role in the energy transition. Because all evaluated portfolios satisfy the applicable clean-energy requirements and emissions limits, selecting $\Gamma$ does not determine whether the CCA meets its decarbonization obligations. Rather, it affects how those obligations are met, including the composition and timing of renewable procurement, dependence on uncertain export revenues, and exposure to future market conditions. Moderate robustness can therefore support a financially resilient transition toward a lower-carbon portfolio without requiring substantial departures from the baseline procurement plan.

Accordingly, the selection of $\Gamma$ should be treated as a planning decision rather than solely as a model parameter. CCAs can use downside financial protection to identify the level of price-risk mitigation obtained from additional robustness, portfolio stability to assess the procurement adjustments required to achieve that protection, and market dependence to evaluate the resulting exposure to external markets. Taken together, the results indicate that moderate uncertainty budgets capture a substantial share of the practical benefits of robustness, whereas larger budgets primarily provide additional conservatism with diminishing marginal benefits. The preferred $\Gamma$ can therefore be selected according to the CCA's relative priorities for affordability, procurement stability, market exposure, and progress toward a low-carbon electricity portfolio.

\begin{table}[!htbp]
\centering
\caption{Market dependence under unseen price realizations.
Mean and tail metrics are computed over out-of-sample price scenarios for fixed portfolios planned at different uncertainty budgets $\Gamma$.}
\label{tab:market_dependence}
\begin{tabular}{lccc}
\hline
$\Gamma$ 
& Mean Dependence 
& P95 Dependence 
& CVaR$_{95}$ Dependence \\
\hline
0 
& 0.0327 
& 0.0382 
& 0.0452 \\

8 
& 0.0315 
& 0.0366 
& 0.0427 \\

40 
& 0.0285 
& 0.0310 
& 0.0339 \\
\hline
\end{tabular}
\end{table}

\section{Conclusions and Future Work}
\label{sec:conclusion}

This paper presented a robust optimization framework for long-term procurement planning under wholesale market price uncertainty, with a particular focus on Community Choice Aggregators (CCAs). CCAs play an important role in the energy transition by procuring renewable and clean-energy resources on behalf of local communities while balancing affordability, resource adequacy, and exposure to wholesale markets. By extending the open-source GenX capacity expansion model with a budgeted robust optimization formulation, the proposed framework explicitly hedges against adverse market price realizations while remaining computationally tractable for large-scale Integrated Resource Planning (IRP) studies. Using a California CCA planning case, we evaluated how increasing protection against market price uncertainty influences investment decisions, market participation, and system costs.

The results demonstrate that explicitly representing market price uncertainty can materially change optimal procurement strategies. Rather than simply increasing overall system cost, robust optimization alters the economic value of market participation, particularly by reducing the attractiveness of uncertain future export revenues. As a result, optimal portfolios shift away from export-oriented renewable expansion toward more conservative investment strategies that are less dependent on wholesale market transactions. These findings suggest that deterministic planning may overestimate the long-term value of market exports, leading to procurement strategies that are more exposed to adverse price conditions than intended.

A second key finding is that a substantial share of the benefits of robust planning can be achieved at moderate uncertainty budgets. Although increasing the uncertainty budget produces progressively more conservative solutions, the additional economic value of protection diminishes at higher levels of robustness. The out-of-sample analysis shows that a moderate budget reduces downside cost exposure while requiring only modest changes to the deterministic procurement portfolio. Higher uncertainty budgets provide additional reductions in market dependence and relative tail exposure, but at a higher nominal planning cost. These results provide a practical basis for selecting uncertainty budgets in IRP applications according to the planner's priorities for affordability, portfolio stability, and market exposure, rather than treating the robustness parameter as an arbitrary modeling choice.

The results also clarify how market uncertainty interacts with the CCA's contribution to the energy transition. All evaluated portfolios satisfy the applicable clean-energy requirements and emissions limits, but the uncertainty budget affects how these targets are achieved. Robust planning changes the amount, timing, and composition of renewable and storage procurement and reduces dependence on uncertain revenues from surplus electricity exports. It can therefore help CCAs pursue decarbonization requirements through procurement portfolios that are financially more resilient to changing wholesale market conditions.

More broadly, this work demonstrates that robust optimization can be incorporated into modern capacity expansion models without sacrificing computational tractability or transparency. The proposed framework provides decision-makers with an explicit mechanism for evaluating the trade-off between nominal economic performance and protection against adverse market conditions. Because the methodology is implemented within the open-source GenX platform, it can be adapted to other CCAs, utilities, and long-term planning applications seeking to account more explicitly for uncertain wholesale market conditions.

Overall, this study suggests three practical lessons for long-term electricity procurement. First, market price uncertainty should be explicitly represented when procurement strategies depend on future wholesale market transactions. Second, robust planning primarily reduces dependence on uncertain export revenues rather than substantially changing reliance on market imports, leading to different investment decisions from those obtained under deterministic planning. Finally, moderate uncertainty budgets appear to capture a substantial share of the practical benefits of robustness while avoiding unnecessary conservatism. Robust planning therefore provides CCAs with a transparent approach for balancing affordability, market risk, and progress toward a low-carbon electricity portfolio.

Future work should extend the framework by jointly considering multiple sources of uncertainty, including demand growth, fuel prices, renewable resource availability, and technology costs. Adaptive robust optimization formulations could also be investigated to allow investment and operational decisions to respond as uncertainty is gradually revealed. In addition, integrating robust optimization with modeling to generate alternatives (MGA) \citep{decarolis2016modelling} could identify multiple portfolios with comparable cost and risk performance but different technology compositions, levels of market dependence, and implementation requirements. Such an extension would allow CCAs to explore alternative, uncertainty-protected pathways for meeting their long-term clean-energy and procurement objectives.

\ack{Sample text inserted for demonstration.}

\funding{Sample text inserted for demonstration.}

\roles{Conceptualization, M.J., J.H., and N.P.; data curation, M.J. and J.H.; formal analysis and methodology, M.J. and N.P.; funding acquisition, J.H. and N.P.; project administration and supervision, N.P.; visualization, M.J. and N.P.; writing—original draft preparation, M.J.; writing—review and editing, M.J. and N.P}

\data{The input data, model inputs, modified GenX source code, simulation results, and analysis scripts used in this study are available on Zenodo at \url{https://doi.org/10.5281/zenodo.21896076}. These materials support the robust optimization, sensitivity, price-of-robustness, and out-of-sample analyses and enable reproduction of the figures and tables presented in this paper.}


\bibliographystyle{iopart-num}
\bibliography{references.bib}

\clearpage 
\appendix
\section*{APPENDIX}
\section{Extended GenX Formulation for CCA Integrated Resource Planning}
\label{app:extended_genx}
\setcounter{equation}{0}
\counterwithin{equation}{section}

This appendix presents the components of the GenX formulation needed to represent CCA integrated resource planning, including the underlying capacity expansion objective, the annual planning reserve margin requirement, wholesale market transactions, and emissions accounting for unspecified imports. Standard GenX investment and operational constraints that remain unchanged are not reproduced and are available in the GenX documentation \citep{genx}.

\subsection{Nomenclature}
\label{app:nomenclature}
\vskip 6pt 
\noindent
 \begin{minipage}{0.95\linewidth}
 \footnotesize
\textbf{Sets and Index}
\begin{description}[style=nextline, labelwidth=3.5cm, leftmargin=4cm, itemsep=1pt, topsep=1pt]
  \item[$M$] Set of market zones; index $z$.
  \item[$F$] Set of fuels; index $f$.
  \item[$T$] Set of time steps; index $t$.
  \item[$G$] Set of all resources; index $y$.
  \item[$G_f$] Subset of resources using fuel $f$.
  \item[$S$] Set of planning stages; index $s$.
\end{description}
\end{minipage}

\vskip 10pt 
\noindent
 \begin{minipage}{0.95\linewidth}
 \footnotesize
\textbf{Parameters}
\begin{description}[style=nextline, labelwidth=3.5cm, leftmargin=4cm, itemsep=1pt, topsep=1pt]
  \item [$\pi^{VOM}_{y}$] Variable O\&M cost of technology y (\$/MWh).
  \item [$\pi^{\text{FUEL}}_{y,t}$] Fuel cost of technology $y$ at time $t$ (\$/MWh).
  \item [$\overline{\Omega^{size}_y}$] Unit size of technology y (MW).
  \item [$\rho_{y,t} \in \mathbb{R}_+$] Availability factor for technology $y$ at time $t$.
  \item [$\epsilon^{\text{CO2}}_{f,t}$] Emissions rate of fuel $f$ at time $t$ (tonnes/MWh).

  \item[$REQ_{z,p}^{PRM}$] Required PRM obligation for zone $z$, policy $p$ (MW).
  \item[$\epsilon^{PRM}_{y,z,p}$]	ELCC (fraction) of resource y in zone $z$ for PRM policy $p$.
  \item[$\mathrm{PriceCap}_z$] Penalty cost for PRM violations in zone $z$.
  
  \item [$\overline{\Pi^{\text{Mkt}}_{t}}$] Maximum energy that can be imported from the market at time $t$.
  \item [$\overline{\Theta^{\text{Mkt}}_{t}}$] Maximum energy that can be exported to the market at time $t$.
  \item [$\pi^{\text{Mkt,Imp}}_{t}$] Hourly market import price (\$/MWh).
  \item [$\pi^{\text{Mkt,Exp}}_{t}$] Hourly market export price (\$/MWh).
  \item [$\epsilon_{t}^{\text{Mkt,CO\textsubscript{2}}}$] Emissions rate of electricity purchased from the market (tonnes/MWh).
    
  \item [$\Delta\pi^{\text{Mkt,Imp}}_{t} \ge 0$] Maximum upward deviation from nominal import price at time $t$ (\$/MWh).
  \item [$\Delta\pi^{\text{Mkt,Exp}}_{t} \ge 0$] Maximum downward deviation from nominal export price at time $t$ (\$/MWh).
  \item [$\Gamma$] Budget of uncertainty (fraction of uncertain parameters).
  
  \item[$\omega_t$] Time-step weight.
  \item[$\mu^{\text{OPEX}}_{s}$] Operational expenditure multiplier for stage $s$.
  \item[$DF_{s}$] Discount factor for stage $s$.
\end{description}
\end{minipage}

\vskip 10pt 
\noindent
 \begin{minipage}{0.95\linewidth}
 \footnotesize
\textbf{Variables}
\begin{description}[style=nextline, labelwidth=3.5cm, leftmargin=4cm, itemsep=1pt, topsep=1pt]
 \item [$\Omega_y \in \mathbb{R}_+$] Installed capacity (MW) of technology $y$
 \item [$\Delta_y^{\text{total}} \in \mathbb{R}_+$] Total installed capacity of technology $y$ (MW).
  \item [$\Theta_{y,t} \in \mathbb{R}_+$] Actual energy injected into the grid by technology $y$ at time $t$ (MWh)
  \item [$\Theta^\text{curt}_{y,t} \in \mathbb{R}_+$] Energy curtailed by technology $y$ at time $t$ (MWh)
  \item [$\Theta^{\text{Mkt}}_{z,t} \in \mathbb{R}_+$] Energy exported to the market at time $t$ (MWh)
  \item[$\Pi^{\text{Mkt}}_{z,t} \in \mathbb{R}_+$] Energy imported from the market at time $t$ (MWh)

  \item[$\zeta^{\text{Mkt,Imp}}_{z,t} \in \{0,1\}$] Binary variable indicating market direction (1 = import, 0 = export)
  \item[\ensuremath{s_{z,t}^{\text{Mkt,Imp}} \in [0,1]}] Worst-case activation for market \textbf{import price} in zone $z$ at time $t$
  \item[\ensuremath{s_{z,t}^{\text{Mkt,Exp}} \in [0,1]}] Worst-case activation for market \textbf{export price} at time $t$
  \item[$s^{PRM}_z \geq 0$] Slack variable for PRM violations in zone $z$
\end{description}
\end{minipage}
\vskip 10pt 
\noindent
 \begin{minipage}{0.95\linewidth}
 \footnotesize
\textbf{Dual Variables}
\begin{description}[style=nextline, labelwidth=3.5cm, leftmargin=4cm, itemsep=1pt, topsep=1pt]
 \item [$p,~q^{\text{Mkt,Imp}},~q^{\text{Mkt,Exp}}$] Dual variables associated with the robust reformulation of the objective function and market price uncertainty constraints

\end{description}
\end{minipage}

\subsection{Capacity Expansion Model}
\label{ssec:app_cap_exp}

The baseline capacity expansion model minimizes the net present value (NPV) of total system costs over the planning horizon, capturing annualized investment costs, fixed operation and maintenance (O\&M) expenditures, variable operating costs, and fuel- and emissions-related expenses. The objective function is shown in \autoref{eq:capexp_obj}, where costs incurred in each planning period $s \in S$ are weighted by a discount factor $DF_s$ and an operational expenditure multiplier $\mu^{\text{OPEX}}_s$, together representing the temporal structure of the planning horizon. This formulation corresponds to the standard GenX objective, and the modeling extensions introduced in subsequent sections build directly on this baseline.

\begin{equation}
\label{eq:capexp_obj}
\begin{aligned}
\min~\text{NPV} = 
\sum_{s \in S} DF_{s}\,\mu^{\text{OPEX}}_{s}
\Bigg(
  \sum_{y \in G} 
    \Big[ 
      \pi^{\text{INVEST}}_{y}\,\overline{\Omega}^{\text{size}}_{y}\,\Omega_{y}
      + \pi^{\text{FOM}}_{y}\,\overline{\Omega}^{\text{size}}_{y}\,\Delta^{\text{total}}_{y}
    \Big] \\
\qquad
  + \sum_{y \in G} \sum_{t \in T} 
    \omega_{t}\, \big( \pi^{\text{VOM}}_{y} + \pi^{\text{FUEL}}_{y,t} \big) \,\Theta_{y,t} \Bigg).
\end{aligned}
\end{equation}

Investment costs $\pi^{\text{INVEST}}_y$ represent the annualized capital expenditures associated with new capacity decisions $\Omega_y$, scaled by the resource size parameter $\overline{\Omega}^{\text{size}}_y$. Fixed O\&M costs $\pi^{\text{FOM}}_y$ apply to the total installed capacity $\Delta^{\text{total}}_y$, which includes both existing and newly built capacity and allows for endogenous retirements where applicable. Operational expenditures capture the cost of dispatch across all modeled hours and include variable O\&M costs $\pi^{\text{VOM}}_y$ and fuel costs $\pi^{\text{FUEL}}_{y,t}$. These costs depend on dispatched output $\Theta_{y,t}$ and are scaled by the representative time-step weight $\omega_t$.

The objective is subject to the standard GenX constraints governing investment and retirement decisions, resource operating limits, storage operation, unit commitment, and zonal power balance. Because these constraints are unchanged in this study, they are not reproduced here; the complete standard formulation is provided in the GenX documentation \citep{genx}. The CCA-specific extensions are presented in the following subsections.

\subsection{Planning Reserve Margin}
\label{ssec:app_prm}

In its standard configuration, GenX represents capacity reserve requirements through hourly constraints that simulate the activation of capacity obligations. In contrast, the IRP process defines the Planning Reserve Margin (PRM) as an annual requirement, expressed in terms of the total effective load-carrying capability (ELCC) needed to meet reliability targets. To align the model with IRP practice, we incorporate an annual PRM constraint that ensures each planning zone $z$ procures sufficient ELCC-adjusted capacity.

GenX uses PCAP accounting, under which each resource contributes to the PRM based on its exogenously specified ELCC value $\epsilon^{\text{PRM}}_{y,z,s}$. The PRM requirement for zone $z$ and planning stage $s$ is enforced through the following constraint:
\begin{equation}
\sum_{y \in G_z}
\epsilon^{\text{PRM}}_{y,z,s} \cdot \Delta^{\text{total}}_y
\;+\;
s^{\text{PRM}}_z
\;\ge\;
REQ^{\text{PRM}}_{z,s},
\qquad \forall z \in M,\; s \in S,
\label{eq:prm_constraint}
\end{equation}
where $\Delta^{\text{total}}_y$ denotes the installed capacity of resource $y$, and $s^{\text{PRM}}_{z,s} \ge 0$ is a slack variable that captures any shortfall in meeting the PRM requirement. The PRM requirement is specified separately for each planning stage based on the corresponding resource-adequacy obligation.

Violations of the PRM constraint are discouraged through a penalty term added to the objective function:
\begin{equation}
\sum_{z \in M}\sum_{s \in S}
DF_s\,\mathrm{PriceCap}_{z,s}\,s^{\text{PRM}}_{z,s}.
\label{eq:prm_penalty}
\end{equation}

\subsection{Market Module}
\label{app:market_module}

Resources located outside the CAISO balancing authority area may participate in the wholesale market through dynamic scheduling. Because these resources are not obligated to deliver electricity to CAISO, their contributions are represented as market transactions corresponding to unspecified imports. The contribution of market imports and exports to the objective function is defined in \eqref{eq:market_obj}.

Market transactions are incorporated into the zonal power balance by adding market imports $\Pi^{\text{Mkt}}_{z,t}$ as a source of electricity and market exports $\Theta^{\text{Mkt}}_{z,t}$ as an additional use of electricity:

\begin{equation}
\text{ePowerBalance}_{z,t}
+ \Pi^{\text{Mkt}}_{z,t}
- \Theta^{\text{Mkt}}_{z,t}
= 0,
\qquad \forall z \in M,\; t \in T.
\label{eq:app_market_balance}
\end{equation}

Import and export quantities are bounded by their corresponding transaction limits:

\begin{align}
0
\leq \Pi^{\text{Mkt}}_{z,t}
&\leq
\overline{\Pi}^{\text{Mkt}}_{z,t},
&&\forall z \in M,\; t \in T,
\label{eq:app_market_import_bound}
\\
0
\leq \Theta^{\text{Mkt}}_{z,t}
&\leq
\overline{\Theta}^{\text{Mkt}}_{z,t},
&&\forall z \in M,\; t \in T.
\label{eq:app_market_export_bound}
\end{align}

To prevent economically unrealistic arbitrage, such as exporting energy purchased from the market, exports $\Theta^{\text{Mkt}}_{z,t}$ are restricted to be no greater than the total electricity generated within that zone:

\begin{equation}
\Theta^{\text{Mkt}}_{z,t}
\leq
\sum_{y \in G_z} \Theta_{y,t},
\qquad \forall z \in M,\; t \in T.
\label{eq:app_market_export_generation}
\end{equation}

The extended model includes optional binary constraints that can enforce mutually exclusive imports and exports. These constraints were not used in the experiments presented in this study. Under the assumed market price structure and export restrictions, simultaneous imports and exports did not occur in the resulting solutions. Therefore, all reported scenarios were solved as linear programs.

To maintain consistency with greenhouse gas accounting requirements for California IRP portfolios, emissions associated with unspecified imports are assigned to the importing zone \citep{CPUC_IRP_Inputs_2025}. Market-related CO\textsubscript{2} emissions $\epsilon^{\text{Mkt,CO2}}_{z,t}$ are calculated as

\begin{equation}
E^{\text{Mkt,CO2}}_{z,t}
=
\epsilon^{\text{Mkt,CO2}}_{z,t}
\Pi^{\text{Mkt}}_{z,t},
\qquad
\forall z \in M,\; t \in T,
\label{eq:app_market_emissions}
\end{equation}

The resulting market emissions $E^{\text{Mkt,CO2}}_{z,t}$ are added to the zonal emissions balance and count toward the applicable emissions limit. This treatment ensures that imported electricity is subject to the same emissions accounting framework as in-state generation and does not receive an artificial advantage in the planning model.

\section{Supplementary Material}
\label{app:sm}
\setcounter{figure}{0}
\setcounter{table}{0}
\counterwithin{figure}{section}
\counterwithin{table}{section}

\renewcommand{\thefigure}{\Alph{section}.\arabic{figure}}
\renewcommand{\thetable}{\Alph{section}.\arabic{table}}

\begin{figure}[htbp!]
  \centering
  \includegraphics[width=0.95\textwidth]{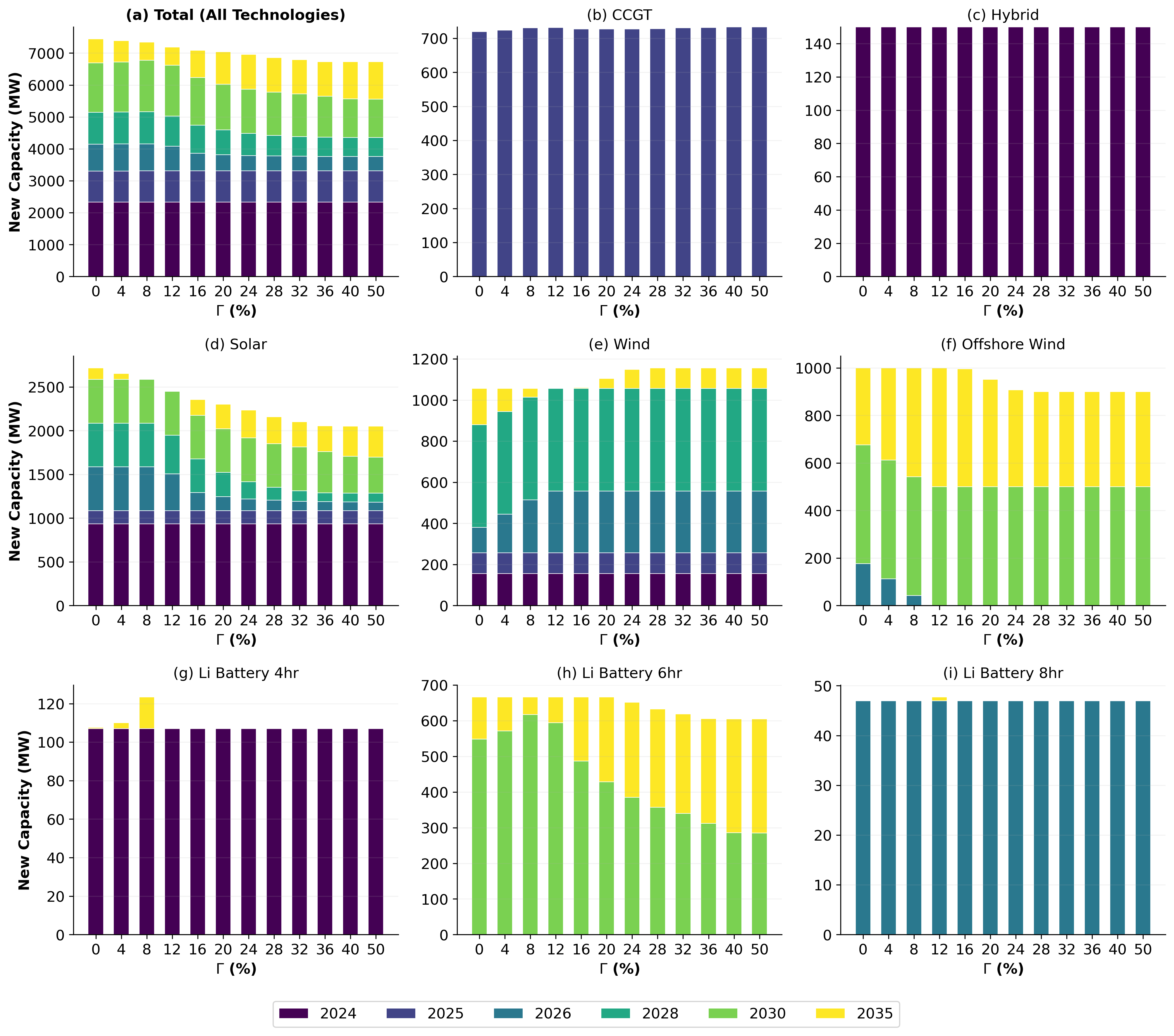}
  \caption{Capacity expansion by technology under increasing market price uncertainty. Panel (a) shows total installed capacity across all technologies, while panels (b)--(i) show selected resource-specific capacity trajectories as a function of the uncertainty budget $\Gamma$. 
Stacked colors indicate the timing of capacity additions across planning years, with 2024 representing existing capacity.}
  \label{fig:baseline_capacity_by_gamma}
\end{figure}

\end{document}